\documentclass{aa}  

\usepackage{graphicx}
\usepackage{txfonts}
\usepackage{amsmath}
\usepackage{lipsum}
\begin{document}

   \title{Nonlinear dynamo models using quasi-biennial oscillations constrained by sunspot area data}
   
   \titlerunning{Nonlinear dynamos using QBOs}
   \authorrunning{Inceoglu et al.}
   \author{F. Inceoglu
          \inst{1,2,3},
          R. Simoniello\inst{4},
          R. Arlt\inst{5}
          \and
          M. Rempel\inst{6}}

   \institute{
    $^{1}$ Department of Engineering, Aarhus University, Aarhus University, Ny Munkegade 120, DK-8000 Aarhus C, Denmark  \\
   email: \url{fadil@eng.au.dk} \\
   $^{2}$ Department of Geoscience, Aarhus University, H{\o}egh-Guldbergs Gade 2, DK-8000 Aarhus C, Denmark\\              
   $^{3}$ Stellar Astrophysics Centre, Department of Physics and Astronomy, Aarhus University, Ny Munkegade 120, DK-8000 Aarhus C, Denmark\\
   $^{4}$ Geneva Observatory, University of Geneva, Geneva, Switzerland \\                 
   $^{5}$ Leibniz-Institut f{\"u}r Astrophysik Potsdam, An der Sternwarte 16, 14482, Potsdam, Germany \\
   $^{6}$ High Altitude Observatory, National Center for Atmospheric Research, Boulder, P.O. Box 3000, Boulder, CO 80307, USA \\
             }
   \date{Received ?; accepted ?}

 
  \abstract
   {Solar magnetic activity exhibits variations with periods between 1.5--4 years, the so-called quasi-biennial oscillations (QBOs), in addition to the well-known 11-year Schwabe cycles. Solar dynamo is thought to be the responsible mechanism for generation of the QBOs.}
   {In this work, we analyse sunspot areas to investigate the spatial and temporal behaviour of the QBO signal and study the responsible physical mechanisms using simulations from fully nonlinear mean-field flux-transport dynamos.}
   {We investigated the behaviour of the QBOs in the sunspot area data in full disk, and northern and southern hemispheres, using wavelet and Fourier analyses. We also ran solar dynamos with two different approaches to generating a poloidal field from an existing toroidal field, Babcock-Leighton and turbulent $\alpha$ mechanisms. We then studied the simulated magnetic field strengths as well as meridional circulation and differential rotation rates using the same methods.}
   {The results from the sunspot areas show that the QBOs are present in the full disk and hemispheric sunspot areas and they show slightly different spatial and temporal behaviours, indicating a slightly decoupled solar hemispheres. The QBO signal is generally intermittent and in-phase with the sunspot area data, surfacing when the solar activity is in maximum. The results from the BL-dynamos showed that they are neither capable of generating the slightly decoupled behaviour of solar hemispheres nor can they generate QBO-like signals. The turbulent $\alpha$-dynamos, on the other hand, generated decoupled hemispheres and some QBO-like shorter cycles.}
   {In conclusion, our simulations show that the turbulent $\alpha$-dynamos with the Lorentz force seems more efficient in generating the observed temporal and spatial behaviour of the QBO signal compared with those from the BL-dynamos.}


   \maketitle
%

\section{Introduction} \label{sec:intro}

As a magnetically active star, the Sun shows cyclic variations in its activity levels. The most known of these variation is the 11-year sunspot cycle \citep{1844AN.....21..233S}, which is superimposed on longer-term variations such as $\sim$90-year Gleissberg cycle \citep{1939Obs....62..158G} and $\sim$210-year Suess cycle \citep{Suess1980Raiocarbon}. In addition to these long-term cyclic variations, the Sun also shows variations in its activity, with periods longer than a solar rotation, but considerably shorter than the Schwabe cycle, such as 8-11 months period in the Ca-K plage index, $\sim$150 day period related to the strong magnetic fields on the Sun \citep{1990SoPh..129..165P}, and the Quasi-Biennial Oscillations (QBOs). The QBOs can be identified across from the subsurface layers \citep{2012A&A...539A.135S,2013ApJ...765..100S} to the surface of the Sun \citep{1998ApJ...509L..49B,2012ApJ...749...27V} and to neutron counting rates measured on Earth \citep{2010SoPh..266..173K,2010ApJ...709L...1V}. The QBOs are therefore believed to be a global phenomenon extending from the subsurface layers of the Sun to the Earth via the open solar magnetic field.

Period of the QBO signal in solar activity indices ranges from 1.5 to 4 years \citep{2009A&A...502..981V}, while its amplitude is in-phase with the Schwabe cycle. The QBOs were observed to be an intermittent signal. They attain their highest amplitude during solar cycle maxima and become weaker during solar cycle minima. The QBOs also develop independently in the solar hemispheres \citep{2014SSRv..186..359B}. Based on NSO/Kitt Peak magnetic synoptic maps, \citet{2012ApJ...749...27V} suggested that the QBOs are distributed equally over all latitudes and are associated with poleward magnetic flux transportation from lower solar latitudes during the maximum of a solar cycle and its descending phase. They also suggested that they found an equator-ward drift that takes $\sim$2 years time in the radial and east-west components of the magnetic field that were calculated using the magnetic synoptic maps.

There are several physical mechanisms proposed that could cause the observed QBOs; flip-flop cycles, which is defined as the 180$^{\circ}$ shift of the active longitudes with largest active regions \citep{2003A&A...405.1121B}, spatiotemporal fragmentation from differences in temporal variations in the radial profile of the rotation rates \citep{2013ApJ...765..100S}, instability of magnetic Rossby waves in the tachocline \citep{2010ApJ...724L..95Z}, and tachocline nonlinear oscillations, where periodically varying energy exchange takes place between the Rossby waves and differential rotation and the present toroidal field \citep{2018ApJ...853..144D}. Additionally, a secondary dynamo working in the subsurface layers as the mechanism behind the QBOs is also proposed \citep{1998ApJ...509L..49B,2010ApJ...718L..19F}, as the helioseismic observations revealed a subsurface rotational shear layer extending to a depth of 5\% of the solar surface \citep{1998ApJ...505..390S}.

Results from the semi-global direct numerical simulations (DNS) show that in addition to the dominant dynamo mode with a longer period, which is generated in the region where the equator-ward migration of the field is observed near surface, there is a weaker pole-ward migrating dynamo mode with s shorter period working in below the top of the domain \citep{2016A&A...589A..56K}. On the other hand, global magnetohydrodynamic (MHD) simulations \citep{2018ApJ...863...35S} revealed that there are two types of cycles. These cycles are nonlinearly coupled in the turbulent convective envelopes. The longer cycle originates in the bottom of the convection zone, while the shorter cycle is generated in the subsurface layers of the domain.

The solar dynamo is the physical mechanism, where the motion of the electrically conducting fluid, the solar plasma, can support a self-excited dynamo that maintains the global magnetic field in the convective envelope of the Sun against ohmic dissipation \citep{1955ApJ...122..293P,1955ApJ...121..491P}, that generates and governs the spatiotemporal evolution of the magnetic activity of the Sun. A large-scale magnetic field can be generated via rotating, stratified, and electrically conducting turbulence. This process is generally referred to as the $\alpha$-effect and it converts kinetic energy of the convection into magnetic energy. The precise nature of these non-dissipative turbulence effects and the $\alpha$-effect in particular, are still under discussion. 

There are two basic processes involved in exciting an oscillatory self-sustaining dynamo: (i) generation of a toroidal field from a pre-existing poloidal field, and (ii) re-generation of the poloidal field from the generated toroidal field. The shearing of any poloidal field by the solar differential rotation can generate the toroidal magnetic field. This process is known as the $\Omega$-effect. As for re-generating the poloidal field from the toroidal one, two of the most promising mechanisms can be described as (i) the effect of rotating stratified turbulence, where helical twisting of the toroidal field lines by the Coriolis force generates a poloidal field (turbulent $\alpha$-effect) \citep{1955ApJ...122..293P,1955ApJ...121..491P,2005PhR...417....1B}, and (ii) the Babcock-Leighton (BL) mechanism \citep{1961ApJ...133..572B,1964ApJ...140.1547L,1969ApJ...156....1L,2014ARA&A..52..251C}. In the BL mechanism, the poloidal field is generated upon the emergence of active regions by uplifting toroidal flux. Diffusion and advection lead to a net transport of flux of predominantly one polarity to the poles, if the active regions are tilted with regards to the east-west direction \citep{1961ApJ...133..572B,1964ApJ...140.1547L,1989Sci...245..712W,1991ApJ...375..761W}.

The inclusion of a poleward surface meridional flow along with an equator-ward deep-seated meridional flow led to the development of the so-called flux transport (FT) dynamo models \citep{1991ApJ...383..431W,1999ApJ...518..508D,2001ApJ...551..576N}. The poleward meridional flow on the surface transports the poloidal field sources to the solar poles and cause the polarity reversal at sunspot maximum \citep{1995A&A...303L..29C,1999ApJ...518..508D}. The meridional flow then penetrates below the base of the convection zone and is responsible for the generation and equator-ward propagation of the bipolar activity structures at low latitudes at the solar surface \citep{1999ApJ...518..508D,2001ApJ...551..576N,2002Sci...296.1671N}. There are several types of FT dynamo models, which produce the poloidal field either via a pure BL-mechanism or a pure $\alpha$-turbulent effect operating in the tachocline, or, alternatively, in the whole convection zone. More recently, dynamo models operating with $\alpha$-turbulence and BL-mechanisms simultaneously as poloidal field sources have also emerged \citep{2001ApJ...559..428D,2013ApJ...779....4B,2014A&A...563A..18P}.

In this study, we investigate the physical mechanisms that could lead to the observed QBOs using solar dynamos. We first analyse sunspot area (SSA) data to decipher the spatiotemporal features of the QBOs in full disk, northern and southern hemispheres (Section~\ref{sec:AN_SSA}). We then describe briefly the fully nonlinear flux transport dynamos with two prevailing approaches to generating a poloidal field from an existing toroidal field; (i) Babcock-Leighton mechanism (Section~\ref{sec:dynBL}), and (ii) turbulent $\alpha$ mechanism (Section~\ref{sec:dyn_modTA}). Following to that, we analyse the simulated data from the two dynamo models to study the possible underlying physics responsible for the spatiotemporal behaviour of the QBO and give the results in Section~\ref{sec:analyses}. We discuss the results and conclude in Section~\ref{sec:dis_conc}.

\section{The QBOs from sunspot areas}
\label{sec:AN_SSA}

To study the spatiotemporal behaviour of the QBO in solar activity, we used publicly available corrected daily sunspot area data\footnote{https://solarscience.msfc.nasa.gov/greenwch.shtml} spanning between 1874--2016. We then calculated the total sunspot areas in full disk (FD), northern hemisphere (NH), and southern hemisphere (SH). In addition, we smoothed the data using moving average with a window length of 6 months to avoid high frequency variations.

\begin{figure}[htb!]
\begin{center}
{\includegraphics[width=3in]{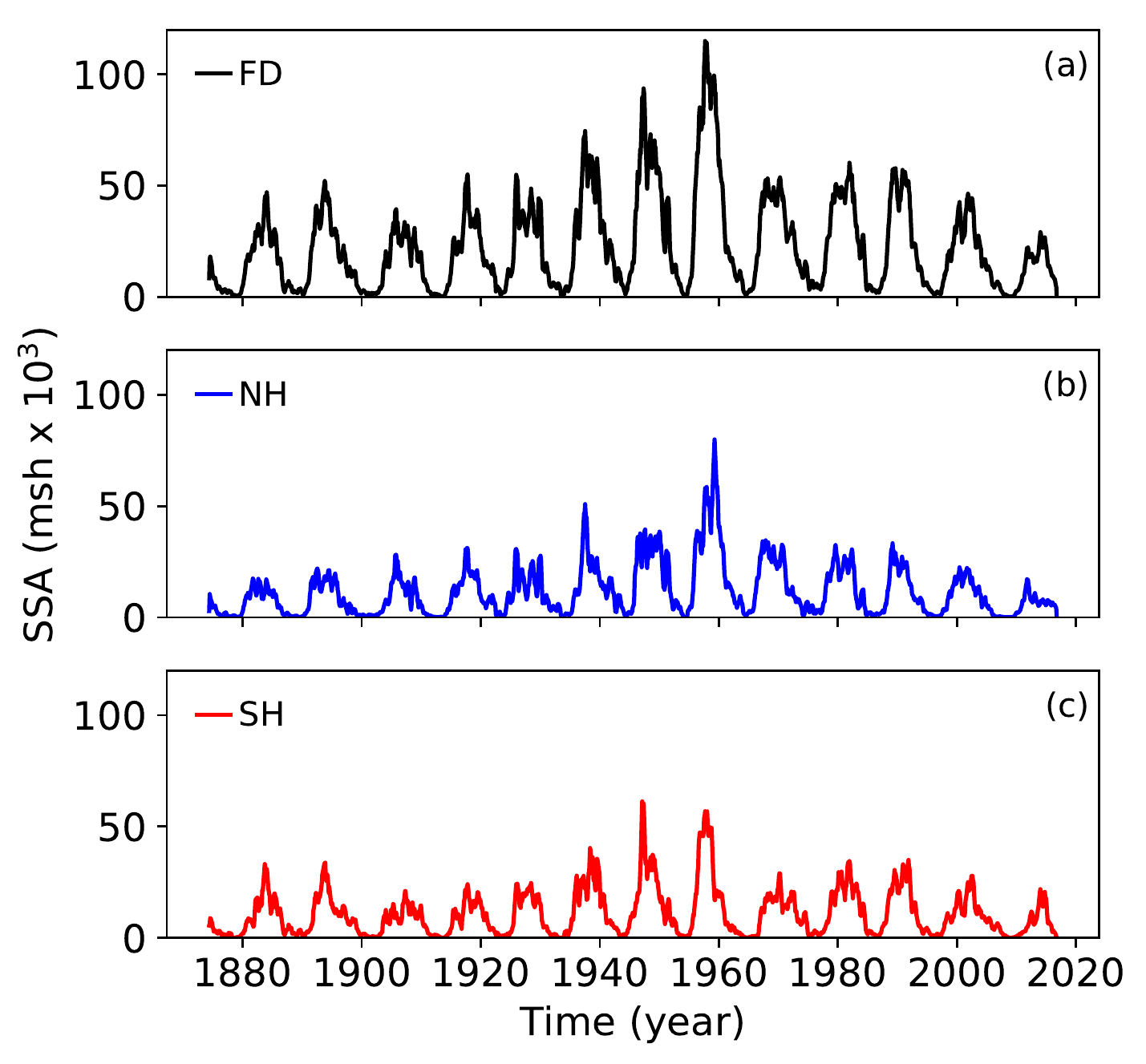}}
\caption{Sunspot areas in millionths of solar hemisphere (a) for full disk (a), northern (b) and southern (c) solar hemispheres.} 
\label{fig:SSA_temp}
\end{center}
\end{figure}

The multi-peak structure during the Schwabe-cycle maxima, which is thought to be related to the QBOs, can be observed in sunspot areas in the FD, NH and SH (Figure~\ref{fig:SSA_temp}). Prior to further analyses, we standardised each SSA data, FD, NH, SH according to their individual mean and standard deviation values.

To further investigate the behaviour of the QBO signal in the SSA data, we first calculated their Fourier power spectral densities (PSDs). To evaluate the significances of the peaks in the calculated Fourier PDSs, we calculated their theoretical Fourier red-noise power spectral densities based on the lag-1 auto-correlation coefficients of each SSA data. We then multiplied each spectrum with the 95$^{th}$ percentile value for $\chi^{2}$ distribution \citep[Equations 16 and 17 in][]{1998BAMS...79...61T}, indicating significance levels of 0.05. The results show that the main period in the SSA data sets is $\sim$11 years corresponding to the Schwabe cycle. Additionally, there are shorter periodicities present in the SSA data below 5 years, which are not the harmonics of the main period of $\sim$11 years and are significant at the 0.05 level. These shorter periods indicate the presence of the QBOs in hemispheric and full disk SSA data (Figure~\ref{fig:SSA_fft}).

\begin{figure}[ht!]
\begin{center}
{\includegraphics[width=3in]{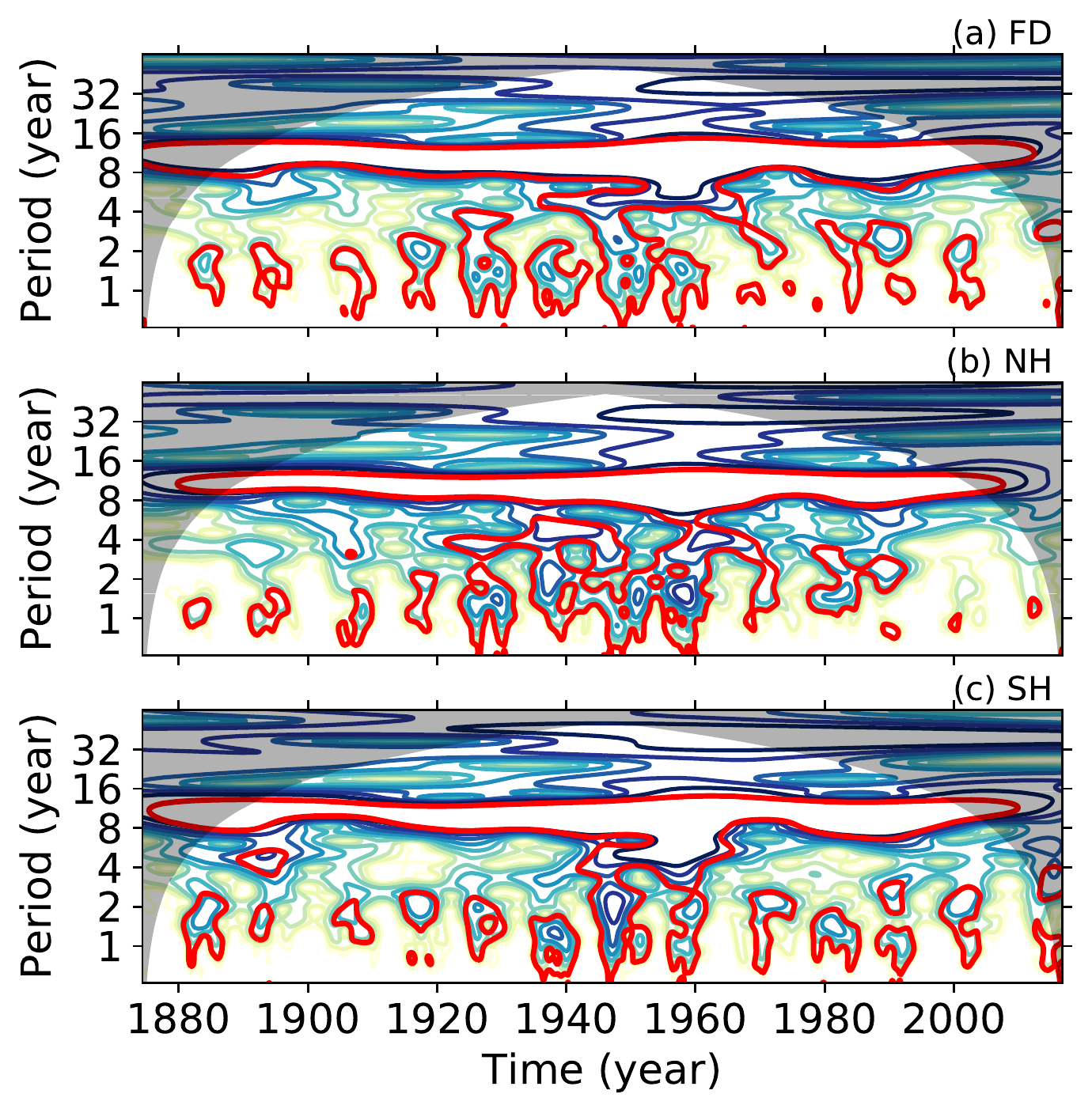}}
\caption{Continuous wavelet spectra of the SSA data in full disk (a), northern hemisphere (b), southern hemisphere (c). Thick red contours indicate the significance level of 0.05.} 
\label{fig:SSA_CWT}
\end{center}
\end{figure}

Additionally, we calculated the continuous wavelet spectra of the each standardised SSA data using the method provided by  \citet{1998BAMS...79...61T} to investigate the temporal behaviour of the QBO signal. The temporal behaviour of the QBO signal in the wavelet spectrum of the full disk SSA data shows in-phase relationship with the Schwabe cycle, where the power of the QBO signal is more prominent during the Schwabe cycle maxima, whereas the QBO signal vanishes during the Schwabe cycle minima. After 1960, the QBO signal during the Schwabe minima vanishes again. After around 2000, the QBO signal is not present in the SSA data (Figure~\ref{fig:SSA_CWT}a).

The QBO signal in the northern hemisphere, similar to the full disk, shows in-phase relationship with the Schwabe cycle, where it is a continuous signal for the period extending from around 1920 to around 1960. It again becomes an intermittent signal in-phase with the Schwabe maxima after around 1960. Different from the full disk data, however, the power of the QBO weakens around 2000 after it becomes longer around 1990 and although weak it comes back around 2015 (Figure~\ref{fig:SSA_CWT}b). The QBO signal in the southern hemisphere, on the other hand, is intermittent and in-phase with the Schwabe cycle maxima throughout the study period. The QBO signal in the southern hemisphere around 2015 seems stronger compared with the one in the northern hemisphere (Figure~\ref{fig:SSA_CWT}c).

\section{The dynamo models}
\label{sec:dyn_mod}

To investigate the physical nature of the QBO signal observed across from the subsurface layers of the Sun to neutron counting rates measured on Earth, we use a FT dynamo model. This model uses the mean field differential rotation and meridional circulation model of \citet{2005ApJ...622.1320R}, which is also coupled with the axisymmetric mean field induction equation \citep{2006ApJ...647..662R}. The $\Lambda$-mechanism in the model, which is responsible for the generation of the mean field differential rotation and meridional circulation is updated following \citet{2005AN....326..379K}. The large-scale flow field alone though does not act as a dynamo, it does amplify and advect magnetic fields. It is modified, on the other hand, by the action of Lorentz forces exerted by the fields generated. 
 
Updating the $\Lambda$-mechanism in the model, meaning using the angular momentum transport equations given in \citet{2005AN....326..379K} led us to use a different set of parameters, which are given in \citet{2017ApJ...848...93I}. The computational domain for our dynamo models extends in latitude from southern to northern pole and in radius from $r=0.65R_{\odot}$ to $0.985R_{\odot}$. For a more detailed information on the dynamo model, we refer the reader to \citet{2005ApJ...622.1320R,2006ApJ...647..662R}.

\begin{figure*}[htb!]
\begin{center}
{\includegraphics[width=5in]{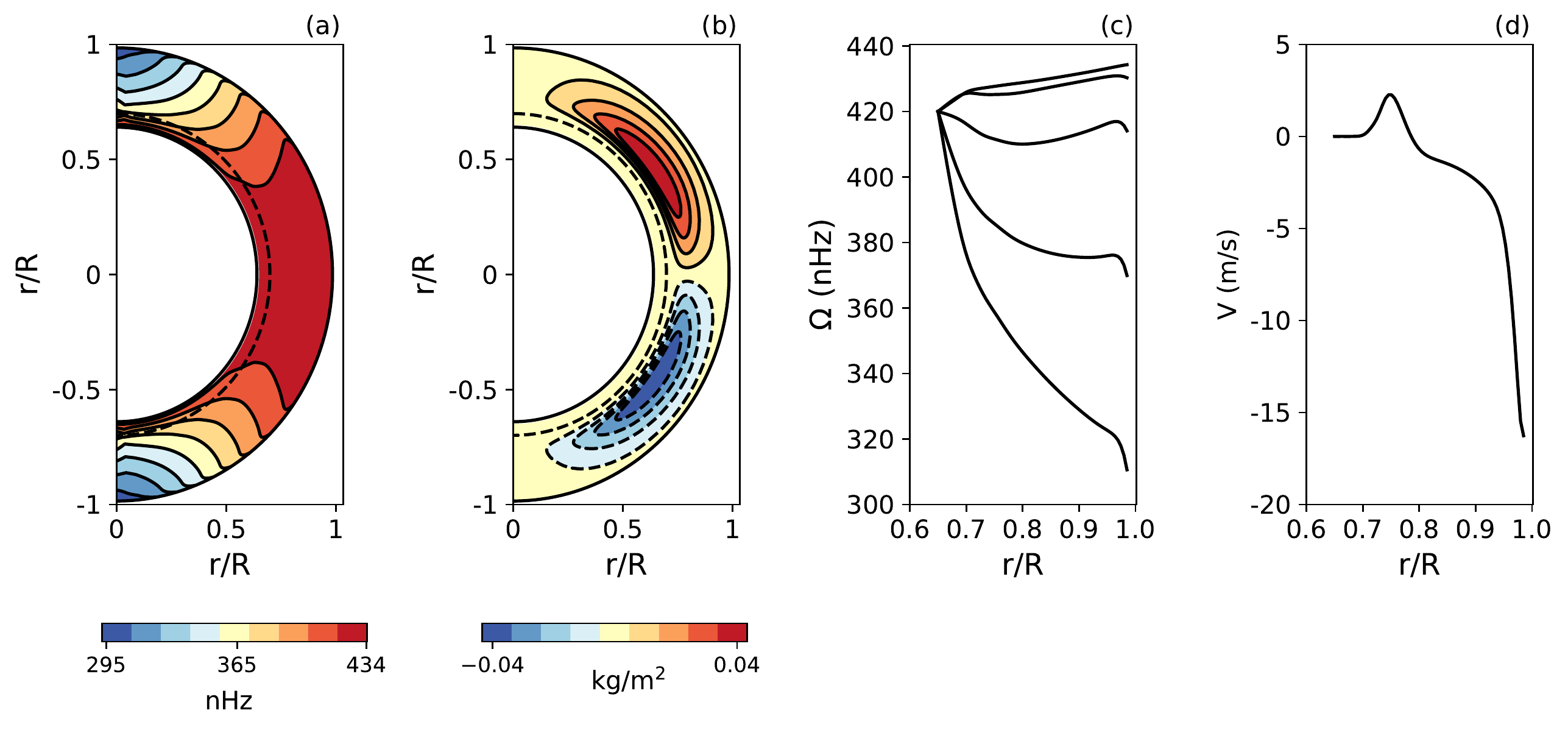}}
\caption{The reference differential rotation (a) and meridional flow in $rsin\theta PSI$, where $PSI$ is the stream function (b) contour plots. Note that the meridional flow contours are normalised according to velocity unit ($(p_{bc}/\varrho_{bc})^{1/2}$). We also show the radial profiles of the reference differential rotation and meridional circulation for the northern solar hemisphere, respectively (c, d). The radial profile of the differential rotation (c) is given at, from top to bottom, 0$^{\circ}$, 20$^{\circ}$, 40$^{\circ}$, 60$^{\circ}$, and 80$^{\circ}$ northern latitudes, while the meridional circulation (d) is given at 45$^{\circ}$ N latitude.} 
\label{fig:MCDF}
\end{center}
\end{figure*}

The computed reference differential rotation and meridional flow radial profiles and their contour plots are shown in Figure~\ref{fig:MCDF}. The differential rotation contours and profile, which is given in units of nHz, show a subsurface shear-layer (Figure~\ref{fig:MCDF}a and c), which is in agreement with the helioseismic observations \citep{2003ARA&A..41..599T,2009LRSP....6....1H}. There is one meridional circulation cell in each hemisphere (shown as $rsin\theta PSI$, where $PSI$ is the stream function, in Figure~\ref{fig:MCDF}b) and the circulation speed on top of the domain at 45$^{\circ}$ N latitude is $\sim$16 m\,s$^{-1}$ (Figure~\ref{fig:MCDF}d).

\subsection{Babcock-Leighton dynamo models}
\label{sec:dynBL}

Dynamo action is accomplished by adding a source term $S(r,\theta,B_{\phi})$, mimicking the BL-effect in the induction equation \cite[see Appendix B and ][for details]{2006ApJ...647..662R}, which is given as:

\begin{eqnarray}
S(r,\theta,B_{\phi})=\alpha_{0}\overline{B}_{\phi, bc}(\theta)f_{\alpha}(r)g_{\alpha}(\theta)
\label{eq:SourceT}
\end{eqnarray}

\noindent with

\begin{eqnarray}
f_{\alpha}(r)=max\bigg[0, 1-\frac{(r-r_{max})^{2}}{d_{\alpha}^{2}}\bigg]
\label{eq:BLalpha}
\end{eqnarray}

\begin{eqnarray}
g_{\alpha}(r)=\frac{(\textrm{sin}\theta)^{4} \, \textrm{cos}\theta}{max\big[(\textrm{sin}\theta)^{4} \, \textrm{cos}\theta\big]}
\label{eq:Galpha}
\end{eqnarray}

\begin{eqnarray}
\overline{B}_{\phi, bc}(\theta)=\int_{r_{min}}^{r_{max}} dr h(r) B_{\phi} (r, \theta)
\label{eq:BLBthetaalpha}
\end{eqnarray}

\noindent where $d_{\alpha}=0.05R_{\odot}$. This confines the poloidal source term above $r=0.935R_{\odot}$, where it peaks at $r_{max}$. The function $h(r)$ is an averaging kernel with $\int_{r_{min}}^{r_{max}} dr \, h(r)=1$. The boundary condition for the model is that the vector potential $A$ and the toroidal field $B_{\phi}$ obey $A=B_{\phi}=0$ and ${\partial A}/{\partial \theta}=B_{\phi}=0$ at the poles and equator, respectively \citep[see Appendix B and ][for details]{2006ApJ...647..662R}.

\begin{figure}[htb!]
\begin{center}
{\includegraphics[width=3in]{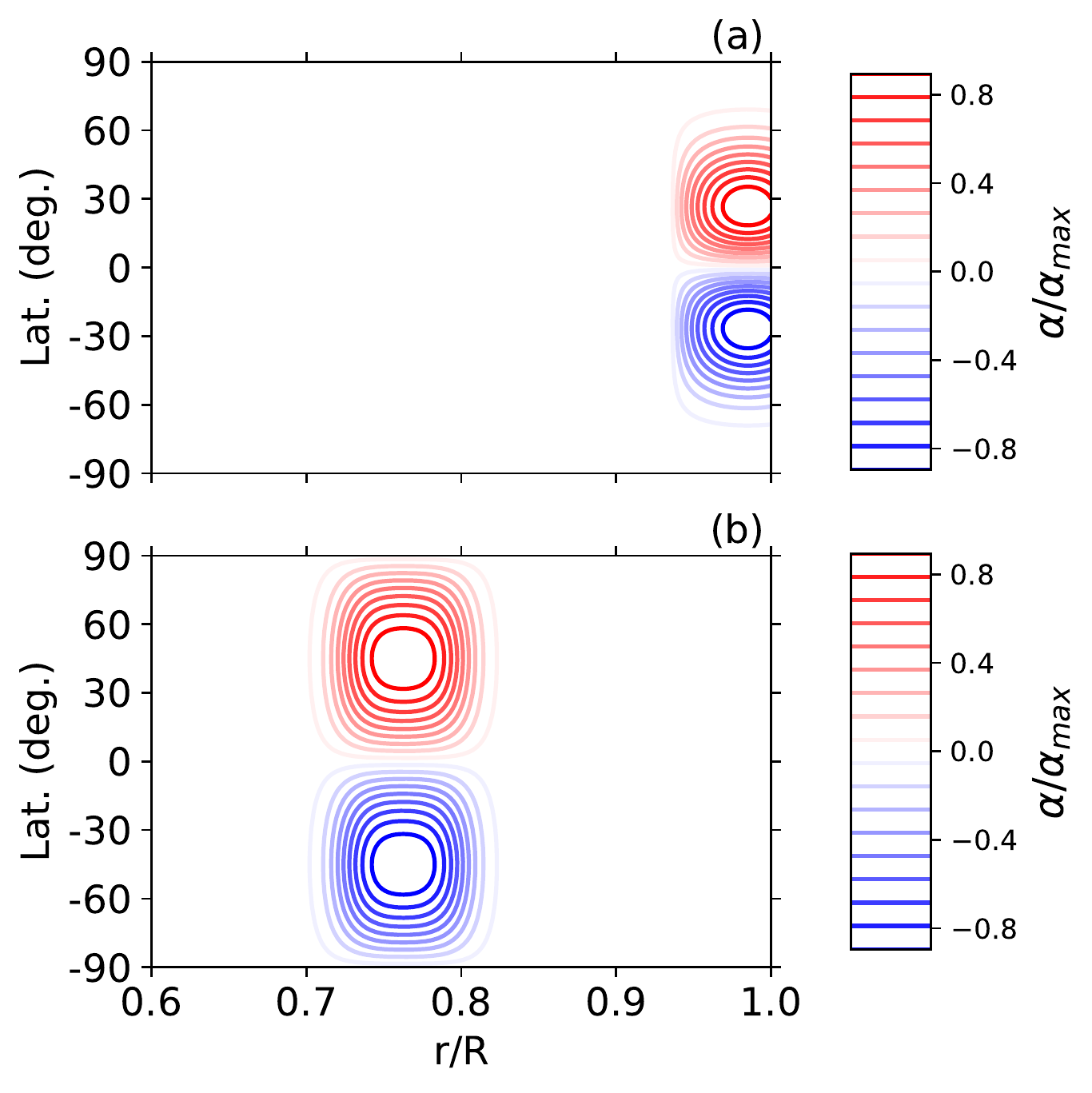}}
\caption{Alpha effect contours for (a) the BL dynamo (Equations~\ref{eq:BLalpha} and \ref{eq:Galpha}) and (b) turbulent $\alpha$-dynamo (Equations~\ref{eq:turbalpha} and \ref{eq:Galpha_turb}) models.} 
\label{fig:Alpha_effect}
\end{center}
\end{figure}

The BL $\alpha$-effect in our simulations is nonlocal, meaning that it operates on the surface (Figure~\ref{fig:Alpha_effect}a) and it is proportional to the toroidal field strength at the base of the convection zone averaged over the interval $[0.71R_{\odot},\, 0.76R_{\odot}]$.

\begin{figure}[htb!]
\begin{center}
{\includegraphics[width=2.5in]{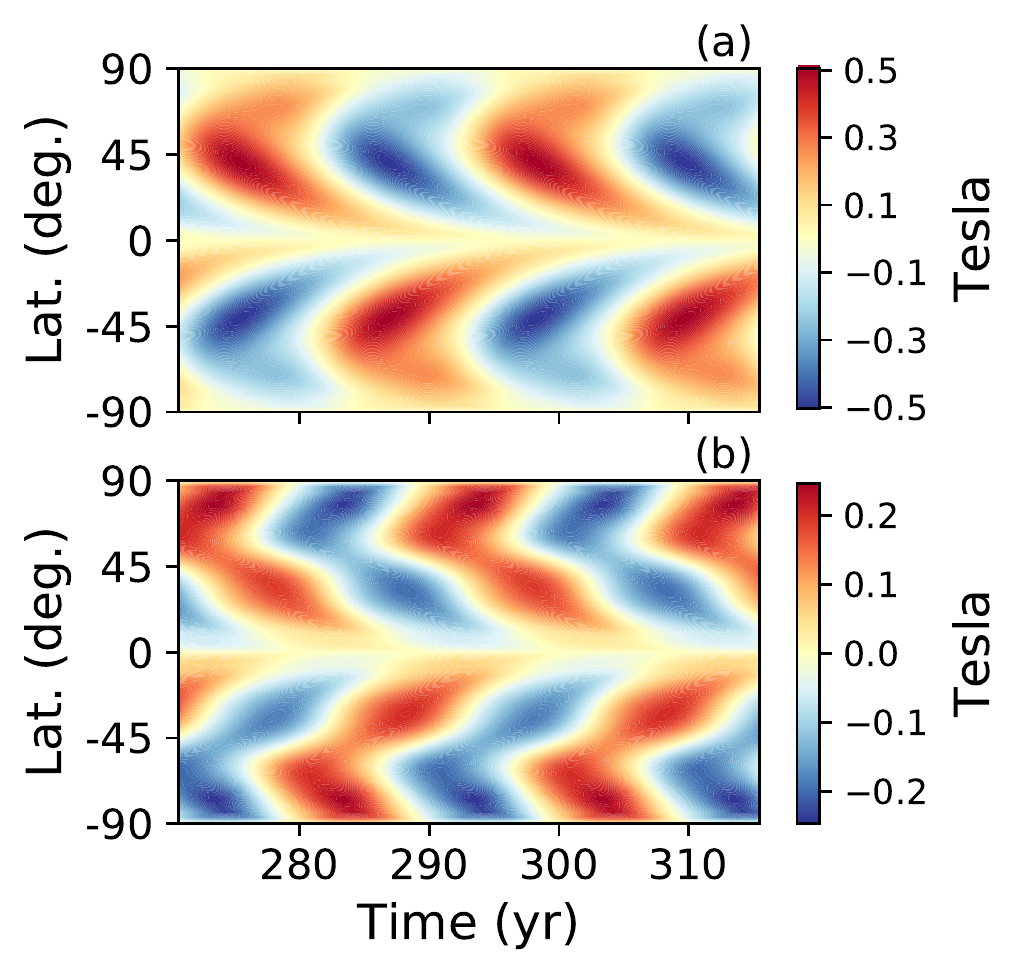}}
\caption{The reference butterfly diagram at 0.71R$_{\odot}$ obtained from (a) the BL dynamo and (b) turbulent $\alpha$-dynamo models.} 
\label{fig:Butter}
\end{center}
\end{figure}

The amplitude of 0.4 ms$^{-1}$ for the $\alpha$-effect coefficient, $\alpha_0$, provided stable anti-symmetric oscillatory behaviour as observed on the Sun. The butterfly diagram computed based on the given configuration at 0.71R$_{\odot}$ shows that the solar cycle starts at around 50$^{\circ}$ latitude at each hemisphere and the magnetic activity propagates equator-ward and poleward (Figure~\ref{fig:Butter}a) as the cycle progresses, resembling the observed sunspot cycles. The average strength of the generated magnetic field in the reference model is around $|B_{\phi}|=3.25$ Tesla, fluctuating between 0.90 and 5.34 Tesla.

\subsection{Turbulent $\alpha$-dynamo models}
\label{sec:dyn_modTA}

The turbulent $\alpha$-effect for the dynamo to generate a poloidal field from a pre-existing toroidal field is taken following \citet{2001ApJ...559..428D} (also see Appendix B for details):

\begin{eqnarray}
f_{\alpha}(r)=\frac{1}{4}\bigg[1+erf\bigg(\frac{r-r_{2}}{d_{2}}\bigg)\bigg]\bigg[1-erf\bigg(\frac{r-r_{3}}{d_{3}}\bigg)\bigg]
\label{eq:turbalpha}
\end{eqnarray}

\noindent with

\begin{eqnarray}
g_{\alpha}(r)=\frac{\textrm{sin}\theta \, \textrm{cos}\theta}{max\big[\textrm{sin}\theta \, \textrm{cos}\theta\big]}
\label{eq:Galpha_turb}
\end{eqnarray}

\noindent where $r_{2}$=0.725R$_{\odot}$, $r_{3}$=0.80R$_{\odot}$, and $d_{2}$=$d_{3}$=0.02R$_{\odot}$. This function constrains the turbulent $\alpha$-effect to a thin layer at the base of the convection zone, just above the tachocline, and to mid-latitudes (Fig~\ref{fig:Alpha_effect}b).

We chose the amplitude of the turbulent $\alpha$-effect coefficient as $\alpha_0$=0.04 ms$^{-1}$, which provided stable anti-symmetric oscillatory solutions. The butterfly diagram calculated based on the given parameters at 0.71R$_{\odot}$ shows that the solar cycles start around 60$^{\circ}$ latitudes at each hemisphere and the magnetic activity propagates towards poles and the equator as the cycles progresses (Figure~\ref{fig:Butter}b). The average strength of the generated magnetic field in the reference turbulent $\alpha$-model is around $|B_{\phi}|=0.96$ Tesla, fluctuating between 0.65 and 1.16 Tesla. As opposed to the BL-dynamo, the turbulent $\alpha$-dynamo exhibits stronger polar branches, whereas equator-ward branches are stronger in  the BL-dynamo (Figure~\ref{fig:Butter}a and b).

\section{Analyses and Results}
\label{sec:analyses}

\subsection{Babcock-Leighton dynamos}
\label{sec:AN_dynBL}

To investigate the physical mechanisms that could lead to the observed relationships in the SSA data sets, we used BL dynamos. To obtain the QBO signals in the BL dynamos, we gradually increased the amplitude of the BL $\alpha$-effect coefficient from 0.4 ms$^{-1}$, which provides stable oscillatory solutions (Figure~\ref{fig:Butter}a), to 1.4 ms$^{-1}$ with 0.1 ms$^{-1}$ increments. Dynamo simulations were not achievable using any larger amplitudes of the BL $\alpha$-effect because of numerical resolution limitations.

\begin{figure}[htb!]
\begin{center}
{\includegraphics[width=2in]{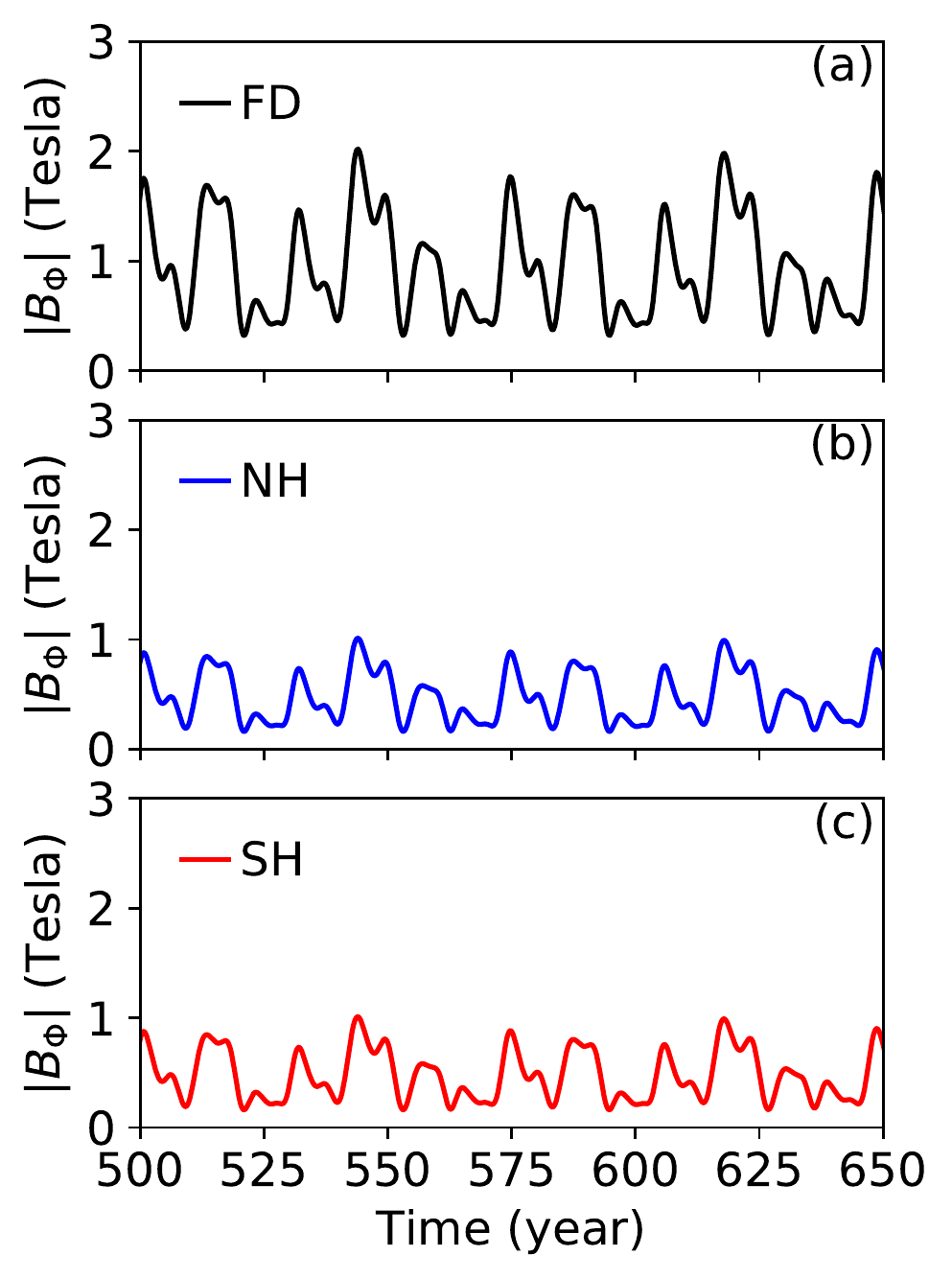}}
\caption{Magnetic field strengths from Babcock-Leighton dynamo for (a) for full disk (black), (b) northern (blue), and (c) southern (red) solar hemispheres.} 
\label{fig:BL_temp}
\end{center}
\end{figure}

We then calculated the total magnetic field strengths at 0.71R$_{\odot}$, which we used for the further analyses. The amplitudes of the total magnetic field strength for the full disk in simulations have decreased from above $|B_{\phi}|=5$ Tesla to around $|B_{\phi}|=2$ Tesla as we increased the amplitude of the BL $\alpha$-effect {bf coefficient} from 0.4 ms$^{-1}$ to 1.4 ms$^{-1}$. This is a direct result of the Lorentz force feedback acting as the saturation mechanism. Here we present the results from the BL dynamo with the $\alpha$-effect coefficient amplitude of 1.4 ms$^{-1}$. 

We calculated the total magnetic field strengths for the northern and southern hemispheres (Figure~\ref{fig:BL_temp}) at 0.71R$_{\odot}$. The BL-dynamo shows similar behaviour between the magnetic field strengths of the northern and southern hemispheres on the contrary to the hemispheric SSA data.

To investigate whether the BL dynamo simulation shows a QBO-like behaviour, we calculated the Fourier power spectral densities using the standardised values of the BL dynamo data for  full disk, and northern and southern hemispheres. Similar to the SSA data, we standardised the data using their individual mean and standard deviation values prior to the Fourier analysis. The results from the Fourier analyses for each data show that the period with the highest power spectral densities are about $\sim$14.8 years for the full disk and the northern and southern hemispheres. There are also signs of periods below 5 years, which can be associated with the QBO, in the Fourier power spectral densities. These periods are close to the 0.05 significance limit for full disk, northern and southern hemispheres (Figure~\ref{fig:BL_fft}).

\begin{figure}[ht!]
\begin{center}
{\includegraphics[width=2.5in]{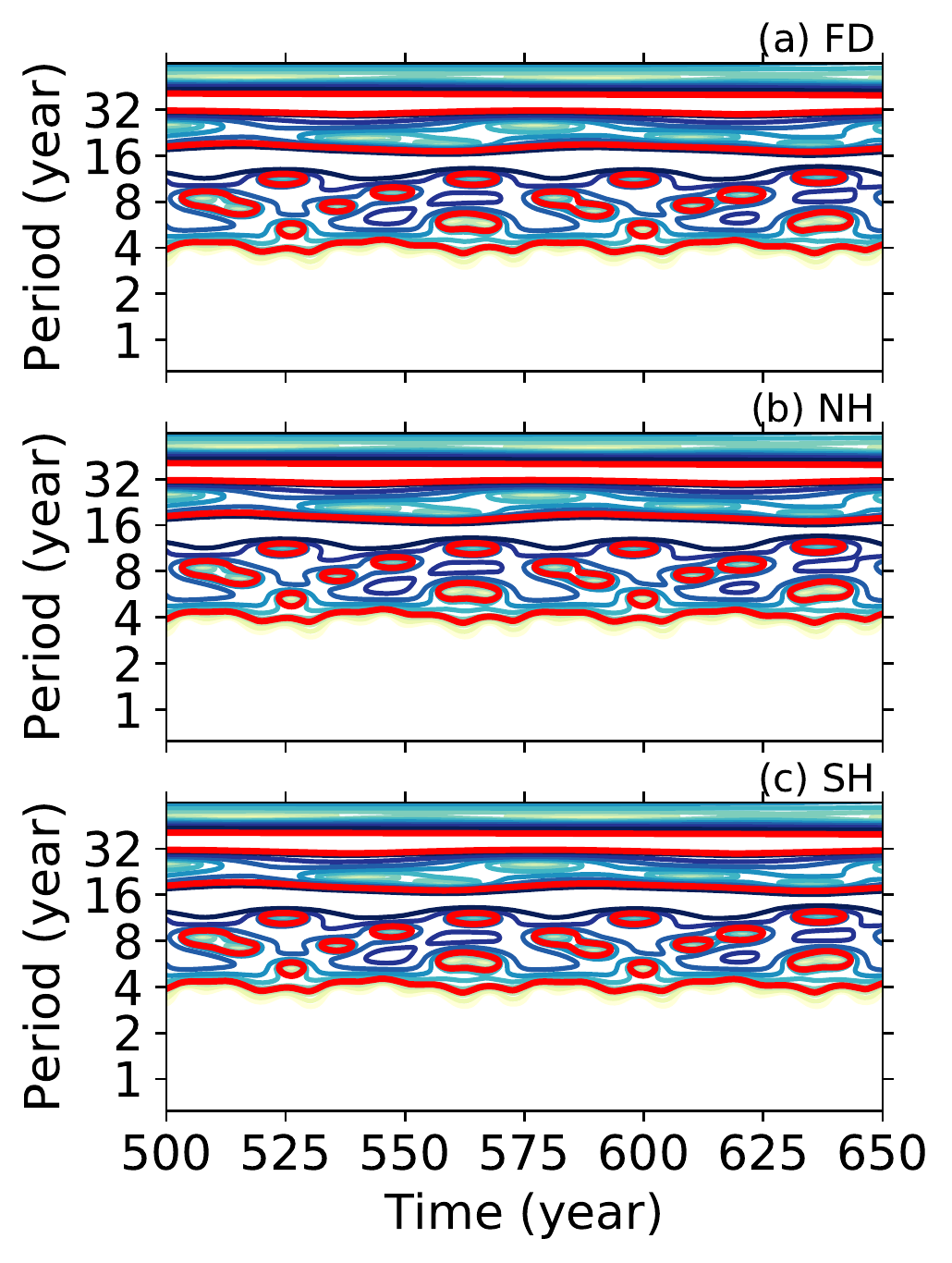}}
\caption{Continuous wavelet spectra of the $|B_{\phi}|$ data from BL-dynamo in full disk (a), northern hemisphere (b), and southern hemisphere (c). Thick red contours indicate the significance level of 0.05.} 
\label{fig:BL_CWT}
\end{center}
\end{figure}

Following to the Fourier analyses, we calculated the continuous wavelet spectra of the simulated data set (Figure~\ref{fig:BL_CWT}). The results from the wavelet analyses show that there are not any periods below about 4 years in the data sets, however there are intermittent signals of a shorter period (SP) between 4 and 8 years in addition to the more continuous longer period (LP) of around 15 years in full disk and in both hemispheres (Figure~\ref{fig:BL_CWT}a and b).

\begin{figure}[htb!]
\begin{center}
{\includegraphics[width=2in]{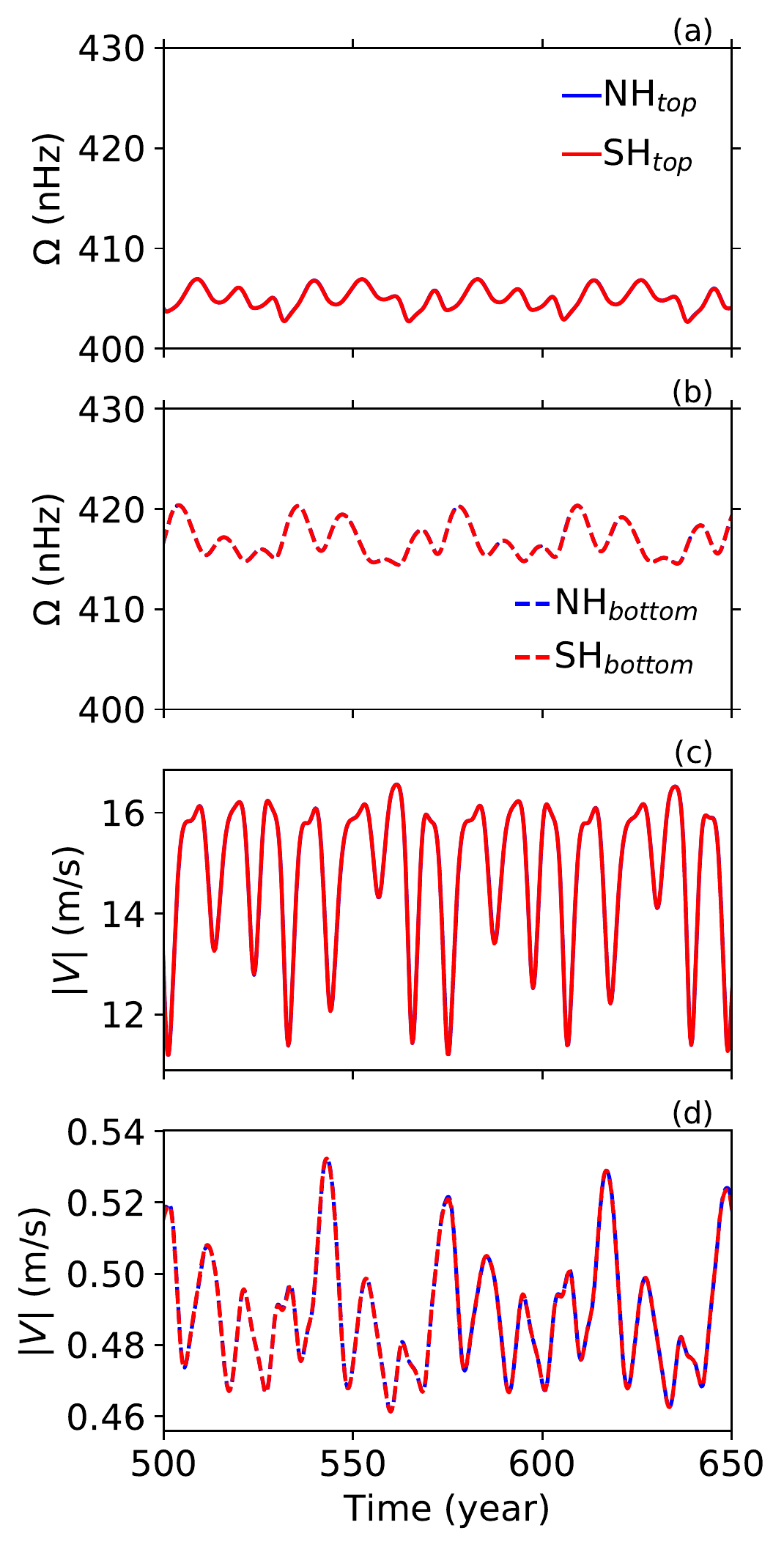}}
\caption{Differential rotation (a, b) and meridional circulation (c, d) rates at the top of the domain 0.985$R_{\odot}$ (lines) and at the bottom of the convection zone 0.71$R_{\odot}$ (dashed lines) at 45$^{\circ}$ latitude N (blue) and S (red) for the BL-dynamo.} 
\label{fig:BL_MCDR_topbot}
\end{center}
\end{figure}
 
To investigate whether the nonlinear interplay between the magnetic and the flow fields might be the reason for the generation of shorter periods observed in the full disk and the hemispheric continuous wavelet spectra (Figure~\ref{fig:BL_CWT}a and b), we analysed meridional circulation and differential rotation values. These values are obtained at 45$^{\circ}$ latitude N and S and at 0.985$R_{\odot}$ and at 0.71$R_{\odot}$ from our BL-dynamo (Figure~\ref{fig:BL_MCDR_topbot}). The differential rotation values at the top of the domain and at the bottom of the convection zone for the both hemispheres are the same and they exhibit small amplitude variations (Figures~\ref{fig:BL_MCDR_topbot}a and b). The meridional circulation values on top of the domain show variations with maximum amplitude of 4 m/s, whereas the maximum amplitude change for the bottom of the convection zone is around 0.04 m/s (Figures~\ref{fig:BL_MCDR_topbot}c and d).

\begin{figure}[htb!]
\begin{center}
{\includegraphics[width=2.5in]{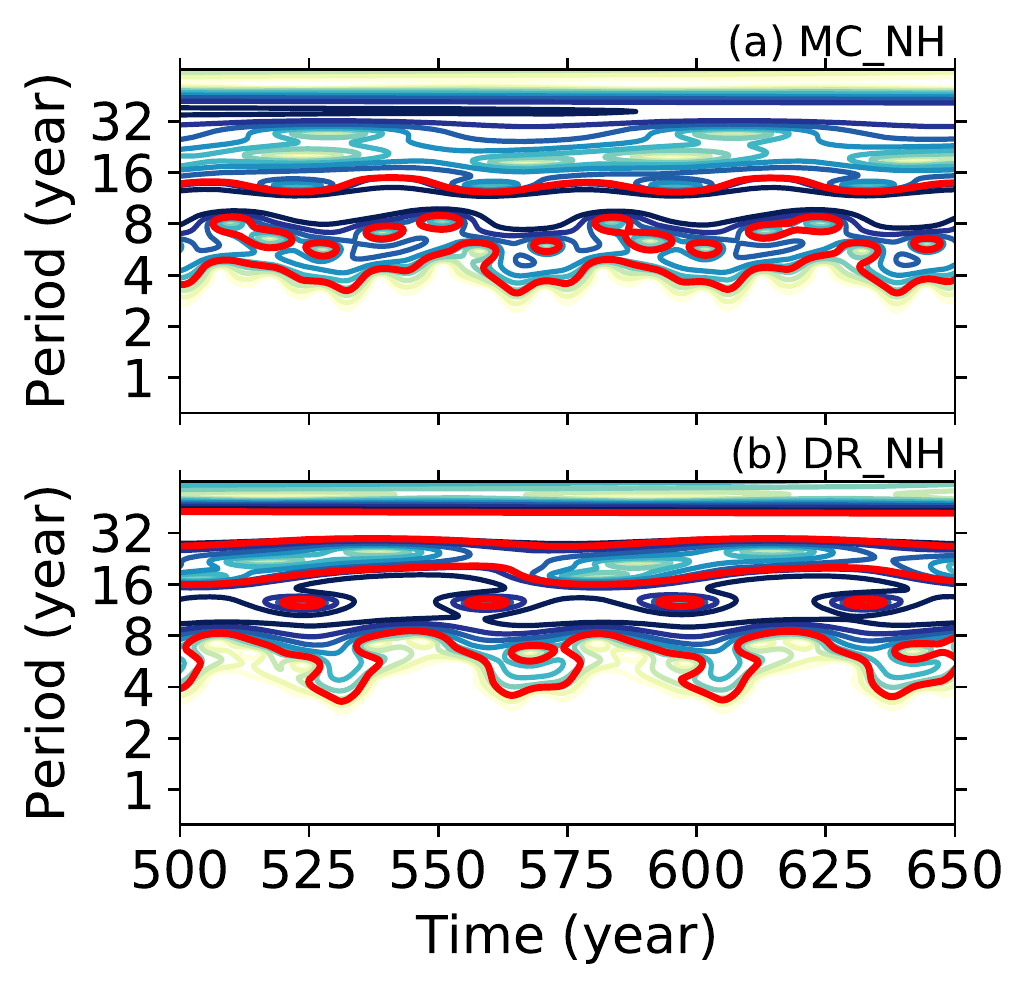}}
\caption{Continuous wavelet spectra of the meridional circulation (a, b) and differential rotation (c, d) values calculated at 45$^{\circ}$ latitude N and S and at 0.985$R_{\odot}$ from the BL-dynamo. Thick red contours indicate the significance level of 0.05.} 
\label{fig:BL_MCDR_wave}
\end{center}
\end{figure}

We then calculated the wavelet spectra of the differential rotation and meridional circulation on top of the domain, as the meridional circulation shows very small amplitude variations at the bottom of the convection zone. Prior to the wavelet analyses, we first standardised the meridional circulation and differential rotation values. We then calculated their wavelet spectra. The meridional circulation in both hemispheres exhibit intermittent longer-term above 8 years and a shorter-term around 4 years variations, latter of which is not present in the differential rotation rates (Figure~\ref{fig:BL_MCDR_wave}).

\subsection{Turbulent $\alpha$-dynamos}
\label{sec:AN_dyn_modTA}

In addition to the BL-dynamos, we also generated the magnetic field strengths using the turbulent $\alpha$-dynamos to study the physical mechanisms leading to the  spatiotemporal relationships observed in the SSA data. To produce the observed features in the SSA data, similar to the BL-dynamos, we gradually increased the amplitude of the turbulent $\alpha$-effect coefficient from 0.04 ms$^{-1}$, which gives the stable oscillatory solutions (Figure~\ref{fig:Butter}b), to 0.2 ms$^{-1}$ with 0.01 ms$^{-1}$ increments. Dynamo simulations were not feasible using any larger amplitudes of the turbulent $\alpha$-effect due to numerical resolution limitations.

The amplitudes of the total magnetic field strength calculated for the full disk in simulations have increased from below $|B_{\phi}|=1.2$ Tesla to below around $|B_{\phi}|=8$ Tesla as the amplitude of the turbulent $\alpha$-effect coefficient increased from 0.04 ms$^{-1}$ to 0.2 ms$^{-1}$. Throughout the different dynamos with increasing turbulent $\alpha$-effect coefficient amplitudes, the regarding simulated butterfly diagrams showed different symmetries from total symmetry to mixed-parity where the anti-symmetry around the equator prevails. Here we present the results from the turbulent $\alpha$-dynamo with the $\alpha$-effect coefficient amplitude of 0.2 ms$^{-1}$ and mixed parity with prevailing anti-symmetry.

\begin{figure}[htb!]
\begin{center}
{\includegraphics[width=3in]{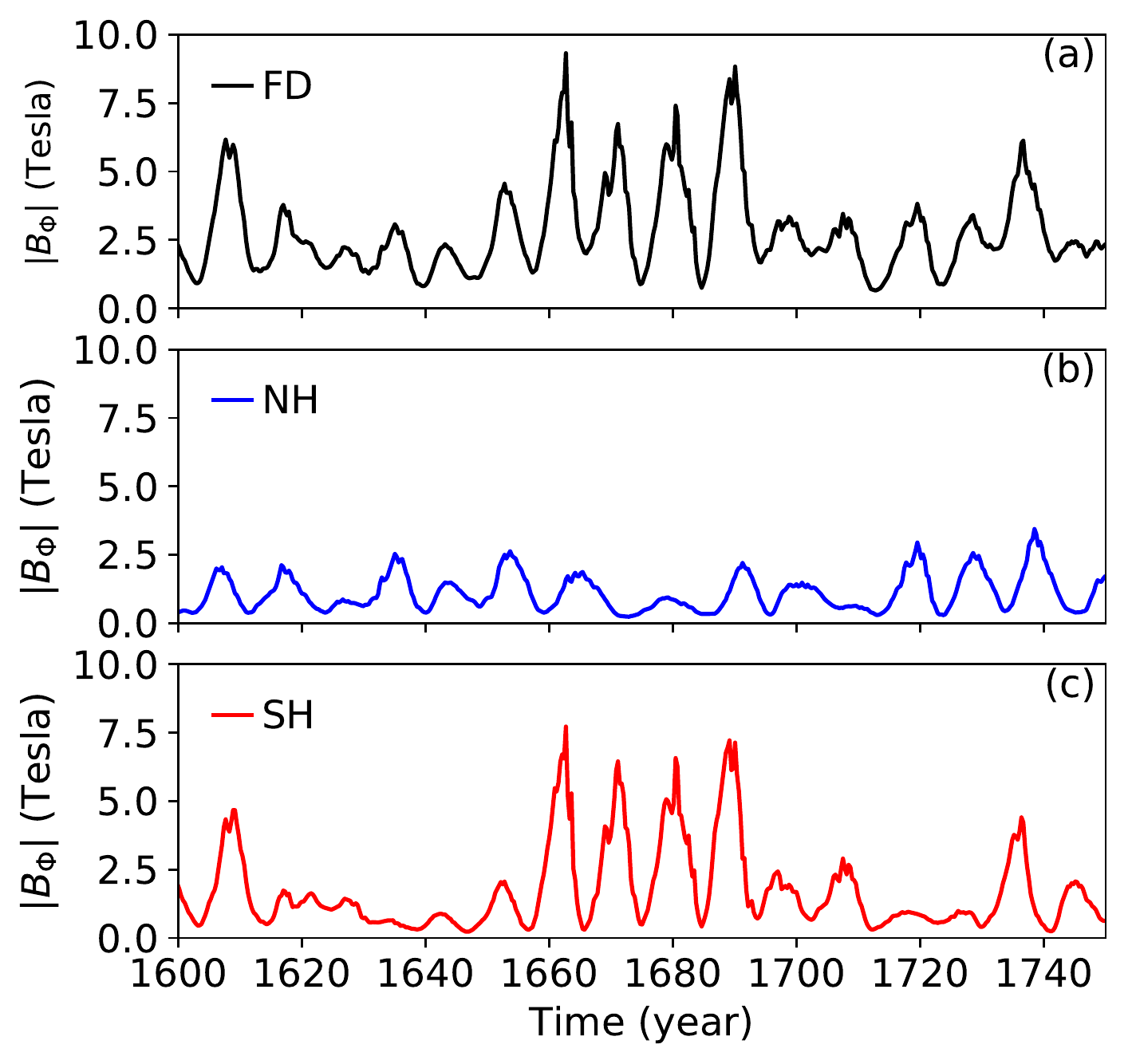}}
\caption{Magnetic field strengths from turbulent $\alpha$-dynamo for (a) for full disk (black), (b) northern (blue), and (c) southern (red) solar hemispheres.} 
\label{fig:DG_temp}
\end{center}
\end{figure}

We calculated the total magnetic field strengths for the northern and southern hemispheres at 0.71R$_{\odot}$ (Figure~\ref{fig:DG_temp}). Different from the SSA data but similar to the BL-dynamo, the magnetic field strengths calculated for the northern and southern hemispheres show a decoupled behaviour, which becomes prominent for the period between around 1650 and 1710.

\begin{figure}[ht!]
\begin{center}
{\includegraphics[width=2.5in]{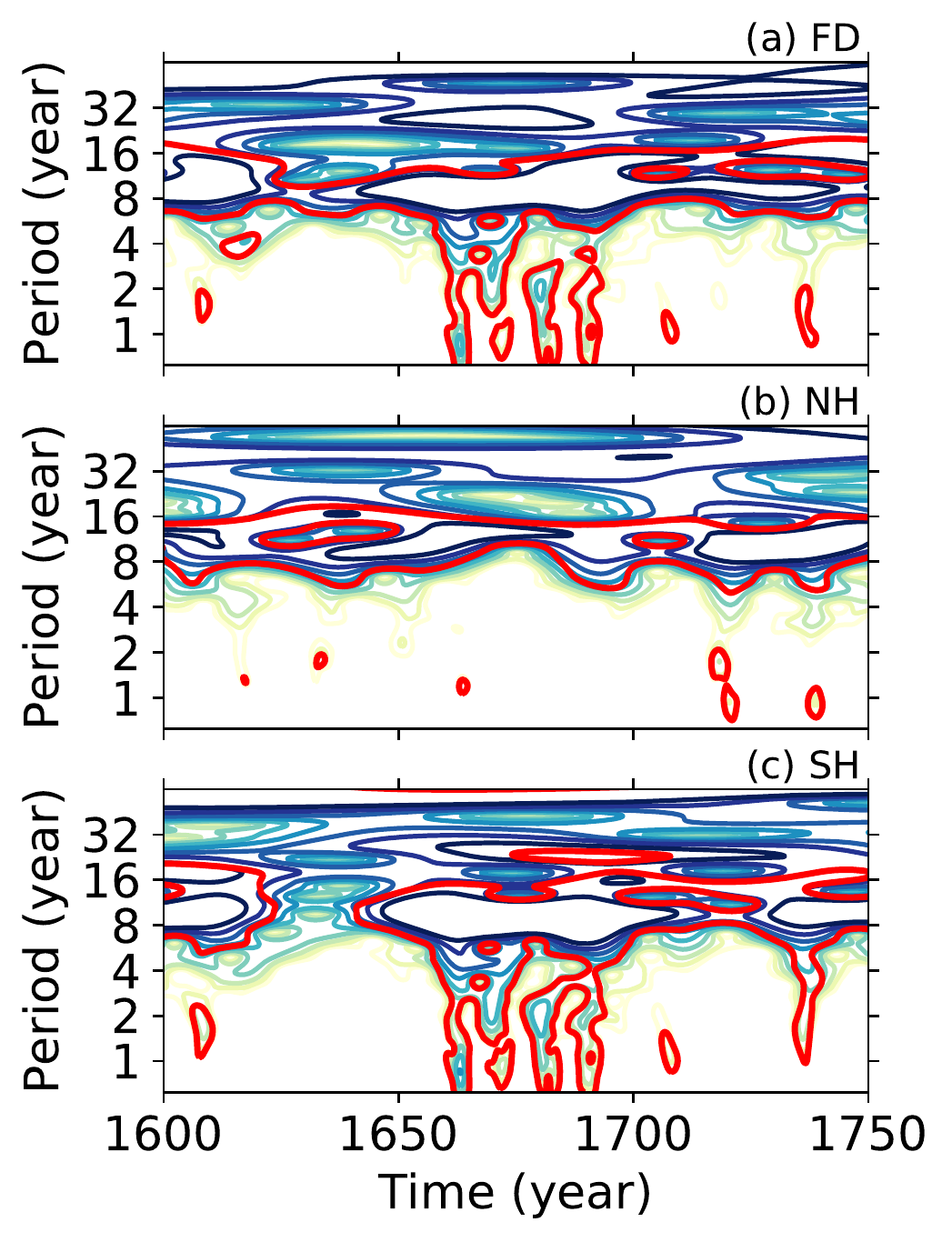}}
\caption{Continuous wavelet spectra of the $|B_{\phi}|$ data from turbulent $\alpha$-dynamo in full disk (a), northern hemisphere (b), and southern hemisphere (c). Thick red contours indicate the significance level of 0.05.} 
\label{fig:DG_CWT}
\end{center}
\end{figure}

To investigate whether the turbulent $\alpha$-dynamo shows a QBO-like behaviour, we calculated the Fourier power spectral densities using the standardised values of the total magnetic field calculated for the full disk, northern and southern hemispheres. Similar to the SSA data, we standardised the simulated data using their individual mean and standard deviation values prior to the Fourier analysis. The results from the Fourier analyses show that the period with the highest power spectral densities are about $\sim$9.5 years across all data together with periods below 5 years that can be associated with the QBO in the Fourier power spectral densities. These periods are statistically significant at the 0.05 level with changing amplitudes across full disk, northern and southern hemispheres (Figure~\ref{fig:DG_fft}). 

In addition to the Fourier analyses, we calculated the continuous wavelet spectra of the simulated data set. The results of the wavelet power spectra show that there are periods around 9.5 year across all data sets as well as periods below 5 years, which can be associated with the QBO signal in the SSA data (Figure~\ref{fig:DG_CWT}). These shorter periods as well as the longer periods are intermittent throughout the simulation period and they exhibit an in-phase variation. The presence of the shorter period is less pronounced in the northern hemisphere than those in the southern hemisphere. The intermittency of both the shorter and longer periods seems to be originating from variations in the total magnetic field strength calculated for the full disk, northern and southern hemispheres, when compared with Figure~\ref{fig:DG_temp}.

\begin{figure}[htb!]
\begin{center}
{\includegraphics[width=2in]{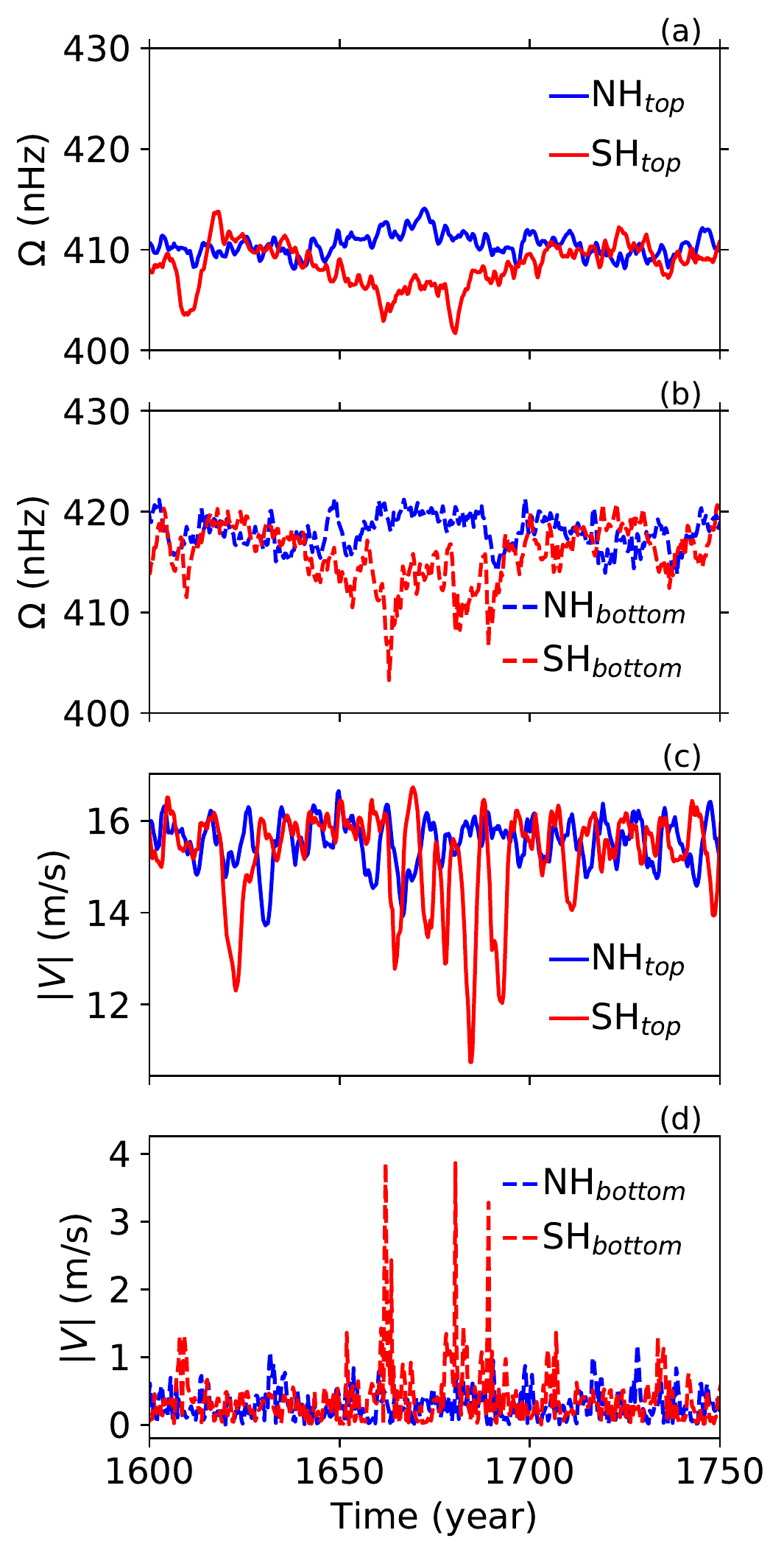}}
\caption{Differential rotation (a, b) and meridional circulation (c, d) rates at the top of the domain 0.985$R_{\odot}$ (lines) and at the bottom of the convection zone 0.71$R_{\odot}$ (dashed lines) at 45$^{\circ}$ latitude N (blue) and S (red) for the turbulent $\alpha$-dynamo.} 
\label{fig:DG_MCDR_topbot}
\end{center}
\end{figure}

To investigate the nonlinear interaction between the magnetic and flow fields as an underlying mechanism to the observed QBO-like signal in our turbulent $\alpha$-dynamo (Figure~\ref{fig:DG_CWT}), we study the variations in the meridional circulation and differential rotation values  at 45$^{\circ}$ latitude N and S and on top of the domain at 0.985$R_{\odot}$, and at the bottom of the convection zone at 0.71$R_{\odot}$ (Figure~\ref{fig:DG_MCDR_topbot}). The differential rotation rates in southern hemisphere both on top of the domain and at the bottom of the convection zone display an anti-correlated relationship with those in the northern hemisphere (Figure~\ref{fig:DG_MCDR_topbot}a and b). The differential rotation rates show their maximum variation of around 15 nHz between the dates 1650 and 1700 at the bottom of the convection zone, while the amplitude of variation in differential rotation rates in the norther hemisphere is around 5 nHz. Similar to the differential rotation values, the meridional circulation speed in the southern hemisphere shows larger variations compared with that in the northern hemisphere. The meridional circulation on top of the domain slows down by 4 m/s in the southern hemisphere, while it experiences episodes with increased speed by more than 3 m/s at the bottom of the convection zone (Figure~\ref{fig:DG_MCDR_topbot}c and d) between 1650 and 1700. Interestingly, this period also shows stronger magnetic field strengths together with QBO-like oscillations only in the southern hemisphere, whereas northern hemisphere mainly shows oscillations with around 10-year period (Figure~\ref{fig:DG_CWT}b and c).

\begin{figure}[htb!]
\begin{center}
{\includegraphics[width=2.5in]{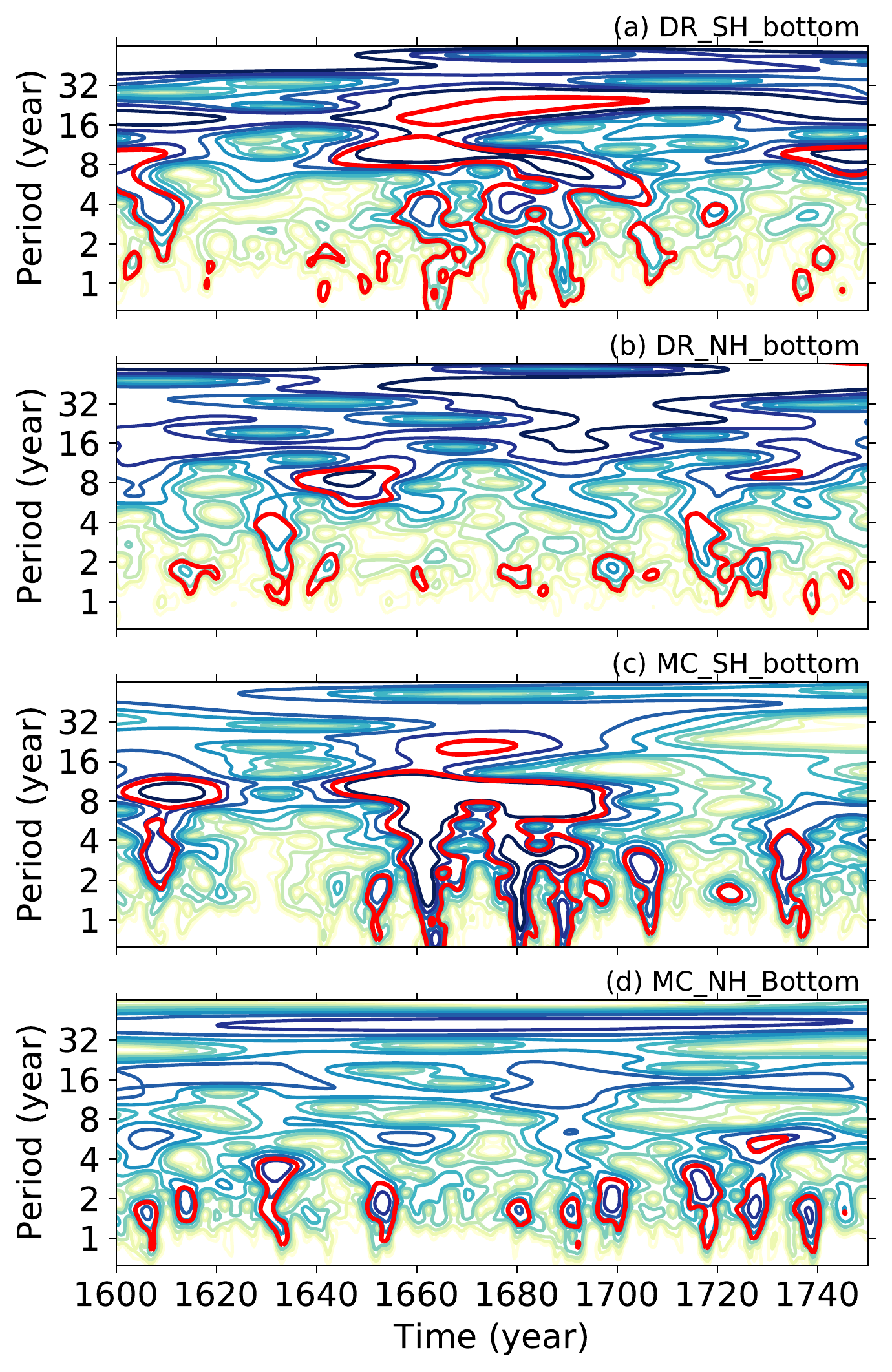}}
\caption{Continuous wavelet spectra of the meridional circulation (a, b) and differential rotation (c, d) values calculated at 45$^{\circ}$ latitude N and S and at 0.71$R_{\odot}$ from the turbulent $\alpha$-dynamo. Thick red contours indicate the significance level of 0.05.} 
\label{fig:DG_MCDR_wave}
\end{center}
\end{figure}

To investigate whether the meridional circulation and differential rotation show any QBO-like signal, we calculated their wavelet power spectra. Prior to the wavelet analyses, we first standardised the meridional circulation and differential rotation values using their individual mean and standard deviation values. The results, which are presented for the bottom of the convection zone show that both the meridional circulation and the differential rotation shows intermittent periods between 1 and 4 years and a longer but statistically not significant at $p<0.05$ level periods (Figure~\ref{fig:DG_MCDR_wave}). A strong and statistically significant periods extending from around 9 years to below 2 years in the southern hemisphere meridional circulation values in around 1650 (Figure~\ref{fig:DG_MCDR_wave}b) can also be observed in the magnetic field strengths in full disk and southern hemisphere (Figure~\ref{fig:DG_CWT}a and c). The intermittency in the QBO-like signals in both the magnetic and flow fields seems to covary (Figures~\ref{fig:DG_CWT} and~\ref{fig:DG_MCDR_wave}).

\section{Discussion and Conclusions}
\label{sec:dis_conc}

The results from the Fourier and wavelet analyses for the sunspot areas in the full disk, and northern and southern hemispheres, reveal that the QBOs are present in sunspot area data with slight differences in temporal and hemispheric behaviours. The observed QBOs generally show an intermittent behaviour having the maximum power during the Schwabe cycle maxima, while the QBOs vanish during the Schwabe cycle minima. The QBO signals in the northern hemisphere becomes considerably less prominent than those observed in the southern hemisphere for the period after around 1990. This feature points a decoupled behaviour of the solar hemispheres, which was also observed for the magnetic energies \citep{2017A&A...601A..51I}. The intermittent behaviour of the QBO was also observed in other solar indices, such as the dipole magnetic moment and open magnetic flux of the Sun \citep{2003ApJ...590.1111W}, global solar magnetic field \citep{2013ApJ...768..189U}, and in the frequencies of the acoustic oscillations \citep{2012MNRAS.420.1405B,2012A&A...539A.135S,2013ApJ...765..100S,2016ApJ...828...41S}. 

We therefore decided to use these observational findings to investigate the physical mechanisms responsible for the generation and temporal and hemispheric evolution of the QBOs. To achieve this, we used fully nonlinear flux transport dynamo models with two prevailing approaches to generating a poloidal field from an existing toroidal field; nonlocal BL- and turbulent $\alpha$-dynamo models. 

The results from the nonlocal BL-dynamo runs show that it generally fails to produce QBO-like shorter cycles and the slightly decoupled behaviour of the solar hemispheres. We also investigated whether the dynamo runs with smaller $\alpha$-effect amplitudes show any QBO-like oscillations. As we increased the BL-effect amplitude from $\alpha_0$=0.4 ms$^{-1}$ up to 1.4 ms$^{-1}$ with 0.1 ms$^{-1}$ increments, the main oscillation that has a period of 11.4 years becomes more perturbed where it shows very small amplitude variations superimposed on the main oscillation. However, none of these runs showed QBO-like oscillations.

The turbulent $\alpha$-dynamo, on the other hand, generated QBO-like intermittent oscillations together with the Schwabe-like cycles as well as the decoupled behaviour of the solar hemispheres. To ensure the QBO-like fluctuations do not result from numerical noise, we also re-ran the dynamo code with doubled spatial and temporal resolution. The results from the high-resolution run showed that the QBO-like oscillations indeed a result of the nonlinear interplay between the magnetic and flow fields. The results from the turbulent $\alpha$-dynamo also show that the meridional flow is stronger impacted by short-term field variability than the differential rotation values (visible around 1700 in Figure~\ref{fig:DG_MCDR_wave}). This is expected since the meridional flow, which is a strongly driven weak flow, has much less kinetic energy and a much shorter response time scale \citep{2005ApJ...631.1286R}. Similar to the BL-dynamos, we also investigated the effect of the amplitude of turbulent $\alpha$-effect coefficient, ranging between 0.04 ms$^{-1}$ and 0.2 ms$^{-1}$. For the weakest coefficient, the magnetic field strength oscillates with a period of 10 years without any perturbations. However, as we increase the amplitude of the turbulent $\alpha$-effect coefficient, the stable oscillation obtained for the amplitude of turbulent $\alpha$-effect coefficient of 0.04 ms$^{-1}$ started to exhibit variations that are superimposed and the amplitudes of these variations increased with the increasing alpha coefficient. After the alpha coefficient of 0.15 ms$^{-1}$, the wavelet analyses starts to show QBO-like oscillations.

The results from the BL- and turbulent $\alpha$-dynamos suggests that results are very sensitive to the detailed structure of the dynamo magnetic field. The shapes of our $\alpha$-effects, both for the turbulent- and BL-$\alpha$, show localisation in latitude and radius to a specific degree so that the simulations show solar-like oscillations with solar-like butterfly diagrams. We tried several shapes for our $\alpha$-effects, especially for the turbulent-$\alpha$, and concluded that the $\alpha$-effect shapes used in our simulations provided the best solar-like solutions. Such shapes are generally not supported by the 3D turbulent numerical simulations \citep{2015ApJ...809..149A,2016AdSpR..58.1522S,2018A&A...609A..51W}.

Our results demonstrated that the nonlocal BL dynamo does not reproduce QBO-like features, while the turbulent $\alpha$-effect at the bottom of the convection zone is more effective in generating QBO-like oscillations. A comparison among the meridional circulation speeds, differential rotation rates, and the magnetic field strengths from the BL (Figures~\ref{fig:BL_temp} and~\ref{fig:BL_MCDR_topbot}) and the turbulent $\alpha$-dynamos (Figures~\ref{fig:DG_temp} and~\ref{fig:DG_MCDR_topbot}) revealed the importance of the differential rotation and meridional circulation rates at the bottom of the convection zone. This region is also very close to where our turbulent $\alpha$-dynamo operates, whereas the nonlocal BL-dynamo $\alpha$-effect operates on the surface. This might indicate that in the presence of the local turbulent $\alpha$-effect, the induced magnetic field will immediately lead to a Lorentz force feed back on the differential rotation and meridional circulation on the bottom of the convection zone. In the case of the nonlocal BL $\alpha$-effect, which operates on the surface, the induced magnetic field has to be transported to the bottom of the convection zone by the meridional circulation and diffusion and then lead to a Lorentz force feed back. This delayed response might be the reason for the absence of QBO-like oscillations in nonlocal BL-dynamos.

Additionally, we investigated the effects of the statistical significance levels of the results from the Fourier and  wavelet analyses, we ran them again with significance levels of 0.10 and 0.01. These runs showed very similar results, having slightly different significance contours, which indicate that our results are robust.

The results from the acoustic mode frequency shifts suggest that the QBOs might be driven in the near surface layers where the ratio of the gas pressure to magnetic pressure, $\beta=(P_{gas}/P_{mag})$ is much smaller \citep{2013ApJ...765..100S}. The helioseismic results have been interpreted either as the presence of a secondary dynamo near the subsurface layers \citep{1998ApJ...509L..49B,2010ApJ...718L..19F} or as the results of a beating between different magnetic configurations \citep{2003A&A...405.1121B,2013ApJ...765..100S}. Further, semi-global DNS and global MHD simulations seem to generate QBO-like shorter cycles in the near-surface layers \citep{2016A&A...589A..56K,2018ApJ...863...35S}. Results from the global MHD simulations also suggest that these two cycles are not necessarily generated by a single dynamo mechanism \citep{2018ApJ...863...35S}, indicating the possibility of a secondary dynamo working in the near surface layer of the Sun. In this region the ratio of the gas pressure to magnetic pressure dramatically decreases, indicating a change in the magnetic field strength or local topology \citep{2013ApJ...765..100S}, and it accommodates a rotational shear-layer \citep{1998ApJ...505..390S}.

In conclusion, our simulations show that the turbulent $\alpha$-dynamos with the Lorentz force seems more efficient in generating the observed temporal and spatial behaviour of the QBO signal compared with those from the nonlocal BL-dynamos. It could also be interesting to investigate whether equator-ward migration from the delay of flux rise to the surface rather than from the deep meridional flow \citep{2018A&A...620A.135F} could lead to stronger QBOs in BL-dynamos. However, to fully understand the specific physical mechanism leading to the presence of QBO-like signals in the turbulent $\alpha$-dynamos, and their absence in the nonlocal BL-dynamos would require a very detailed analyses of the terms in the dynamo equations, such as Lorentz force, shear, advection, and stretching. Additionally, to assess whether a secondary dynamo operating on the surface or in the near-surface layers could generate the observed features in the sunspot areas, a detailed work is necessary. For this work, we plan to use dynamos with turbulent $\alpha$-mechanism in the bottom of the convection zone and the BL-mechanism on the surface as well as two turbulent $\alpha$-dynamos places in the bottom of the convection zone and in the top of the convection zone.

\begin{acknowledgements}

We thank Mausumi Dikpati for her useful comments. We also thank the anonymous reviewer for the comments, which improved the manuscript. Funding for the Stellar Astrophysics Centre is provided by the Danish National Research Foundation (grant agreement no. DNRF106). The National Center for Atmospheric Research is sponsored by the National Science Foundation.

\end{acknowledgements}

\begin{appendix} 
\section{Fourier Analyses of SSA and the BL- and turbulent $\alpha$-dynamos}

\begin{figure}[htb!]
\begin{center}
{\includegraphics[width=2in]{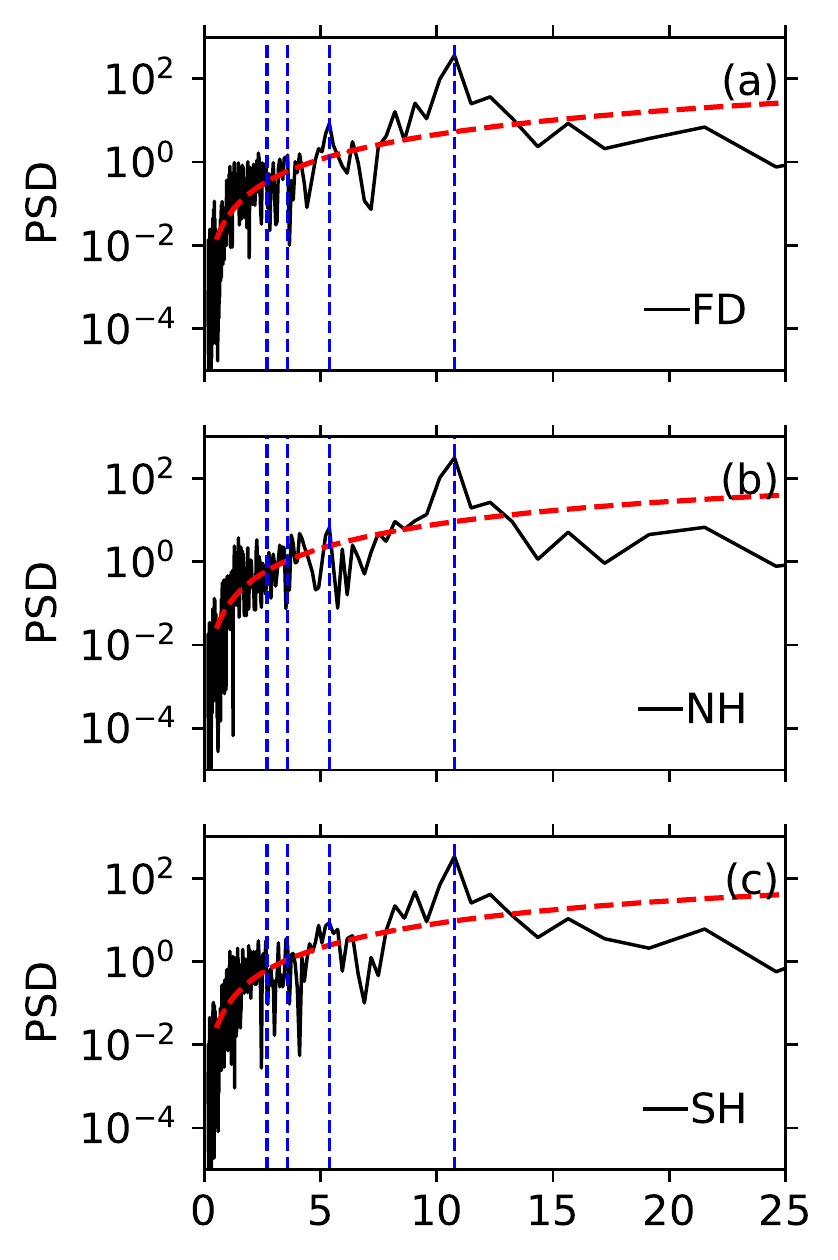}}
\caption{Fourier power spectral densities for the SSA data (black lines), lag-1 auto-correlation Fourier power spectral densities (dashed-red lines) indicating the significance level of 0.05, and the highest period and its first three harmonics (dashed-blue lines) for full disk (a), northern hemisphere (b), and southern hemisphere (c).} 
\label{fig:SSA_fft}
\end{center}
\end{figure}

\begin{figure}[htb!]
\begin{center}
{\includegraphics[width=2in]{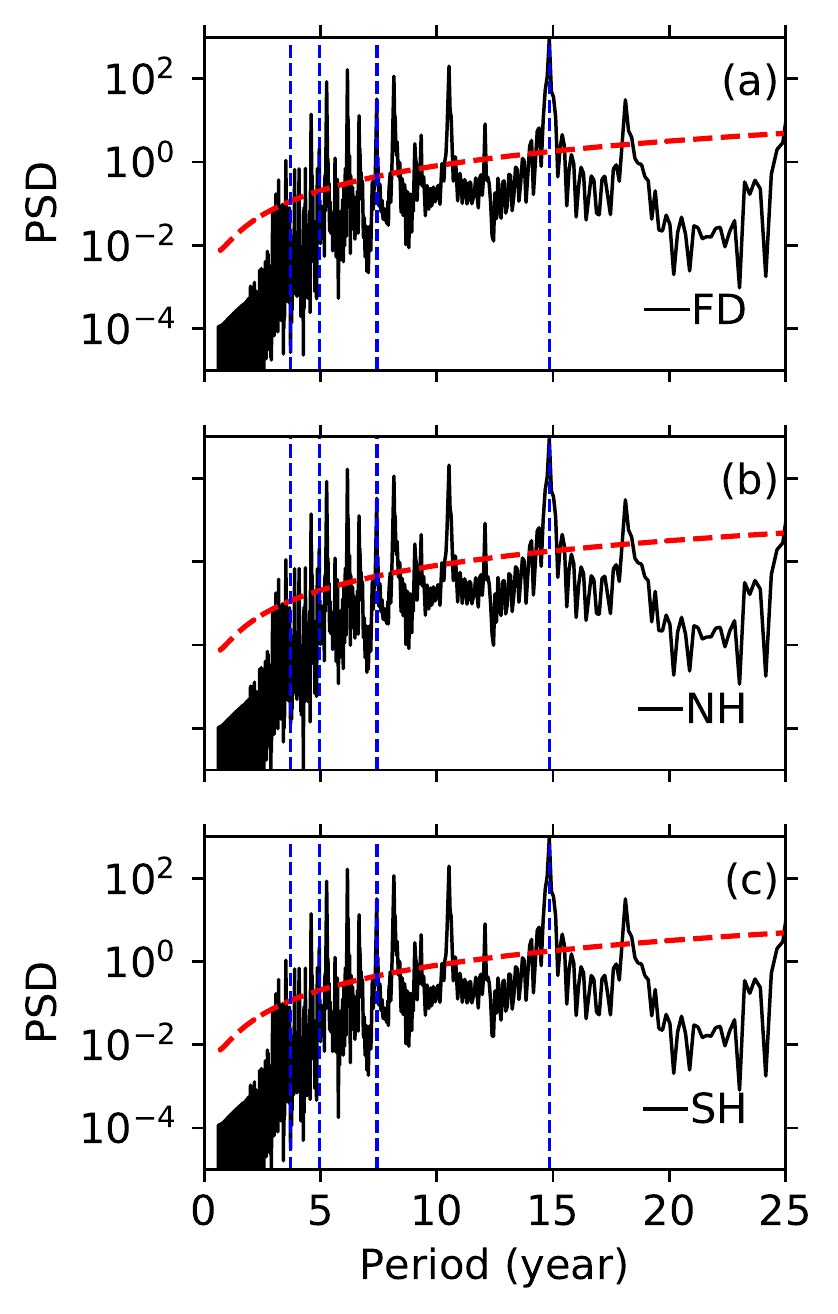}}
\caption{Fourier power spectral densities for the BL-dynamo data (black lines), lag-1 auto-correlation Fourier power spectral densities (dashed-red lines) indicating the significance level of 0.05, and the highest period and its first three harmonics (dashed-blue lines) for full disk (a), northern hemisphere (b), and southern hemisphere (c).} 
\label{fig:BL_fft}
\end{center}
\end{figure}

\begin{figure}[htb!]
\begin{center}
{\includegraphics[width=2in]{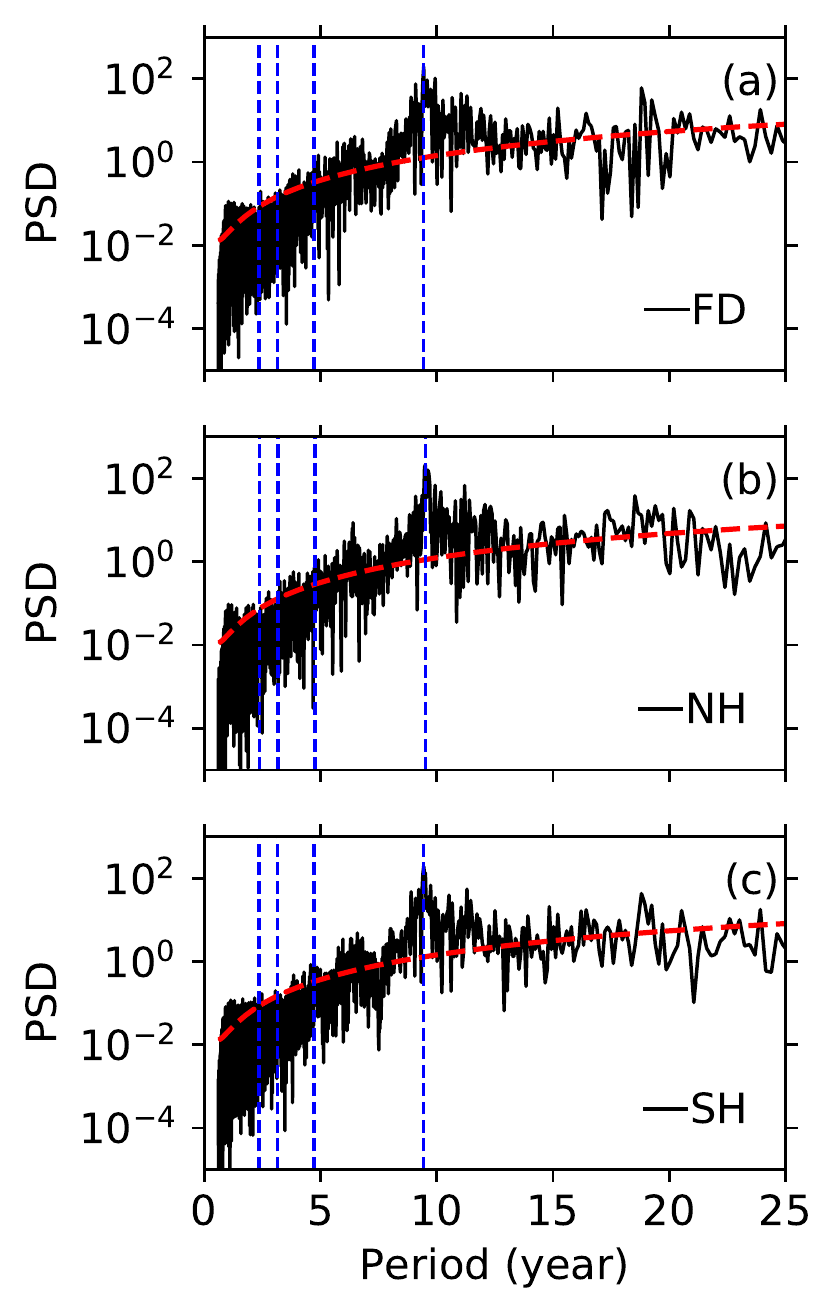}}
\caption{Fourier power spectral densities for the turbulent $\alpha$-dynamo data (black lines), lag-1 auto-correlation Fourier power spectral densities (dashed-red lines) indicating the significance level of 0.05, and the highest period and its first three harmonics (dashed-blue lines) for full disk (a), northern hemisphere (b), and southern hemisphere (c).} 
\label{fig:DG_fft}
\end{center}
\end{figure}

\section{Dynamo Equations}

The following equations are mostly a reproduction from \citet{2005ApJ...622.1320R,2005AN....326..379K,2006ApJ...647..662R}. The basic assumptions made for generation of the differential rotation and meridional circulation in the solar convection zone are as follows.

\begin{itemize}
\item Axisymmetry and a spherically symmetric reference state.
\item All processes on the convective scale are parameterised, which lead to turbulent viscosity, turbulent heat conductivity, and turbulent angular momentum transport.
\item The equations can be linearised under the assumption that $\varrho_{1} << \varrho_{0}$ and $p_{1} << p_{0}$ and spherically symmetric reference state, where $\varrho_{0}$ and $p_{0}$ are the reference state values and $\varrho_{1}$ and $p_{1}$ are the perturbations caused by the differential rotation. Here we solve axisymmetric hydrodynamic equations that are fully compressible and linearised.
\item The entropy equation includes only perturbations from differential rotation and hence the reference state is assumed to be in an energy flux balance. It is also assumed that the effect of convection in the convection and overshoot regions on entropy perturbations associated with the differential rotation is purely diffusive.
\item The tachocline at the base of the convection zone is forced by a uniform rotation boundary condition at $r=0.65 R_{\odot}$.
\end{itemize}

\begin{eqnarray}
\frac{\partial \varrho_{1}}{\partial t} = - \frac{1}{r^2} \frac{\partial}{\partial r} (r^2 \varv_r \varrho_{0}) - \frac{1}{r\, \textrm{sin}\theta} \frac{\partial}{\partial \theta}(\textrm{sin}\theta\, \varv_\theta\, \varrho_{0})\,,
\end{eqnarray}

\begin{eqnarray}
\frac{\partial \varv_{r}}{\partial t} &=& - \varv_{r} \frac{\partial \varv_r}{\partial r} - \frac{\varv_\theta}{r} \frac{\partial \varv_r}{\partial \theta} + \frac{\varv^2_\theta}{r} - \frac{1}{\varrho_0} \bigg[\varrho_1 g(r) + \frac{\partial p_1}{\partial r}\bigg] \\ \nonumber
&&+ (2 \Omega_0 \Omega_1 + \Omega_1^2)\, r\, sin^2 \theta + \frac{F_r}{\varrho_0}\,,
\end{eqnarray}

\begin{eqnarray}
\frac{\partial \varv_{\theta}}{\partial t} &=& - \varv_{r} \frac{\partial \varv_\theta}{\partial r} - \frac{\varv_\theta}{r} \frac{\partial \varv_\theta}{\partial \theta} - \frac{\varv_r \, \varv_\theta}{r} - \frac{1}{\varrho_0} \frac{1}{r} \frac{\partial p_1 }{\partial \theta}  \\ \nonumber
&&+ (2 \Omega_0 \Omega_1 + \Omega_1^2)\, r\, \textrm{sin}\theta\, \textrm{cos}\theta + \frac{F_\theta}{\varrho_0}\,,
\end{eqnarray}

\begin{eqnarray}
\frac{\partial \Omega_{1}}{\partial t} &=& - \frac{\varv_r}{r^2} \frac{\partial}{\partial r} \, \bigg[r^2 (\Omega_0+ \Omega_1)\bigg] \\ \nonumber
&&- \frac{\varv_\theta^2}{r\,\textrm{sin}^2\theta} \frac{\partial}{\partial \theta}\bigg[\textrm{sin}^2\theta \,(\Omega_0 + \Omega_1)\bigg] + \frac{F_\phi}{\varrho_o\,r\,\textrm{sin}\theta}\,,
\end{eqnarray}

\begin{eqnarray}
\frac{\partial s_1}{\partial t} &=& -\varv_r\, \frac{\partial s_1}{\partial r} - \frac{\varv_\theta}{r} \frac{\partial s_1}{\partial \theta} + \varv_r \frac{\gamma \delta}{H_p} + \frac{\gamma - 1}{p_0} Q \\ \nonumber
&& + \frac{1}{\varrho_0 T_0}\,\textrm{div}(\kappa_t\,\varrho_0\,T_0\,\nabla\,s_1)\,,
\label{eq:Annexeq5}
\end{eqnarray}

\noindent where

\begin{eqnarray}
p_1=p_0 \bigg(\gamma\, \frac{\varrho_1}{\varrho_0} + s_1 \bigg)\,,
\end{eqnarray}

\begin{eqnarray}
H_p= \frac{p_0}{\varrho_0 \, g} \,,
\end{eqnarray}

\begin{eqnarray}
\frac{ds_0}{dr} = -\frac{\gamma\delta}{H_p} \,.
\end{eqnarray}

\noindent The last term in Equation~B.5 describes the turbulent diffusion of entropy perturbation within the convection zone, where $\kappa_t$ represents the turbulent thermal diffusivity.

The viscous force $\boldsymbol{F}$ is given as follows.

\begin{eqnarray}
F_r &=&\frac{1}{r^2} \frac{\partial}{\partial r} (r^2 R_{rr}) + \frac{1}{r\,\textrm{sin}\theta} \frac{\partial}{\partial\theta} (\textrm{sin}\theta \, R_{\theta r}) \\ \nonumber
&& - \frac{R_{\theta\theta} + R_{\phi\phi}}{r} \,,
\end{eqnarray}

\begin{eqnarray}
F_\theta &=&\frac{1}{r^2} \frac{\partial}{\partial r} (r^2 R_{r\theta}) + \frac{1}{r\,\textrm{sin}\theta} \frac{\partial}{\partial\theta} (\textrm{sin}\theta \, R_{\theta\theta}) \\ \nonumber
&& + \frac{R_{r\theta} - R_{\phi\phi}cot\theta}{r} \,,
\end{eqnarray}

\begin{eqnarray}
F_\phi &=&\frac{1}{r^2} \frac{\partial}{\partial r} (r^2 R_{r\phi}) + \frac{1}{r\,\textrm{sin}\theta} \frac{\partial}{\partial\theta} (\textrm{sin}\theta \, R_{\theta\phi}) \\ \nonumber
&& + \frac{R_{r\phi} + R_{\theta\phi}cot\theta}{r} \,,
\end{eqnarray}

\noindent with the Reynolds stress tensor

\begin{eqnarray}
R_{ik} &=- \varrho_0 \langle \varv_i^\prime \varv_k^\prime \rangle = \nu_t \varrho_0 \bigg(E_{ik}- \frac{2}{3} \delta_{ik} \textrm{div}\,\boldsymbol{\varv} + \Lambda_{ik} \bigg) \,,
\end{eqnarray}

\noindent where $\Lambda_{ik}$ is the nondifusive Reynolds stress and the deformation tensor, $E_{ik}=\varv_{i;k} + \varv_{k;i}$, is given for the spherical coordinates as follows.

\begin{eqnarray}
E_{rr} = 2 \frac{\partial \varv_r}{\partial r}\,,
\end{eqnarray}

\begin{eqnarray}
E_{\theta\theta} = 2 \frac{1}{r} \frac{\partial \varv_\theta}{\partial \theta} + 2 \frac{\varv_r}{r} \,,
\end{eqnarray}

\begin{eqnarray}
E_{\phi\phi} = \frac{2}{r}\, (\varv_r + \varv_\theta\,cot\theta) \,,
\end{eqnarray}

\begin{eqnarray}
E_{r\theta} = E_{\theta r} = r \frac{\partial}{\partial r} \frac{\varv_\theta}{r} + \frac{1}{r} \frac{\partial \varv_r}{\partial \theta} \,,
\end{eqnarray}

\begin{eqnarray}
E_{r\phi} = E_{\phi r} = r\,\textrm{sin}\theta\, \frac{\partial \Omega_1}{\partial r} \,,
\end{eqnarray}

\begin{eqnarray}
E_{\theta\phi} = E_{\phi\theta} = \textrm{sin}\theta\, \frac{\partial \Omega_1}{\partial r} \,,
\end{eqnarray}

\noindent The energy converted by the Reynolds stress is

\begin{eqnarray}
Q = \sum_{i,k} \frac{1}{2} E_{ik}\,R_{ik} \,.
\end{eqnarray}

\noindent We transform the entropy equation to an equation of the quantity $\varrho_0 T_0 s_1$ that better represents the energy perturbation associated with the entropy perturbation to make importance of the term Q for a stationary solution more apparent. In the case of a stationary solution, we have $\textrm{div}(\varrho_0 \boldsymbol{\varv}) = 0$, which allows us to rewrite the entropy equation in the following form under the assumption of $\mid\delta\mid = \mid \nabla - \nabla_{ad} \mid \ll 1$

\begin{eqnarray}
&&\textrm{div}(\boldsymbol{\varv}\, \varrho_0 T_0 s_1- \kappa_t\, \varrho_0 T_0 \textrm{grad} s_1)  = \\ \nonumber
&&(\gamma - 1) \bigg(\frac{\varrho_0 T_0}{p_0} Q - \varv_r \frac{\varrho_0 T_0 s_1}{\gamma H_p} \bigg)+ \varv_r\gamma\delta\,\frac{\varrho_0 T_0}{H_p} \,.
\end{eqnarray}

For the background $\varrho_0$, $p_0$, and $T_0$, we use an adiabatic hydrostatic stratification, assuming an $\sim r^2$ dependence of the gravitational acceleration given by

\begin{eqnarray}
T_0(r) = T_{bc} \Bigg[1+ \frac{\gamma - 1}{\gamma} \frac{r_{bc}}{H_{bc}} \bigg(\frac{r_{bc}}{r} - 1 \bigg) \Bigg] \,.
\end{eqnarray}

\begin{eqnarray}
p_0(r) = p_{bc} \Bigg[1+ \frac{\gamma - 1}{\gamma} \frac{r_{bc}}{H_{bc}} \bigg(\frac{r_{bc}}{r} - 1 \bigg) \Bigg]^{\gamma/(\gamma - 1)} \,.
\end{eqnarray}

\begin{eqnarray}
\varrho_0(r) = \varrho_{bc} \Bigg[1+ \frac{\gamma - 1}{\gamma} \frac{r_{bc}}{H_{bc}} \bigg(\frac{r_{bc}}{r} - 1 \bigg) \Bigg] ^{1/(\gamma - 1)} \,.
\end{eqnarray}

\begin{eqnarray}
\mathrm{g}(r) = \mathrm{g}_{bc} \bigg(\frac{r}{r_{bc}}^{-2} \bigg)\,. 
\end{eqnarray}

\noindent where $T_{bc}$, $\varrho_{bc}$, and $p_{bc}$ represent temperature, density, and pressure values at the base of the convection zone, $r=r_{bc}$. The term $H_{bc} = p_{bc} / (\varrho_{bc} \mathrm{g}_{bc})$ is the pressure scale height and $\mathrm{g}_{bc}$ is the value of the gravity at $r_{bc}$. In our simulations, we use $r_{bc}=0.71R_{\odot}$, $p_{bc}= 6\times10^{12}$ Pa, $\varrho_{bc}= 200$ kg\,m$^{-3}$, $\mathrm{g}_{bc}=520$ m\,s$^{-2}$, and $R_{\odot}=7\times10^{8}$ km.

For the superadiabaticity $\delta$ we assume the following profile:

\begin{eqnarray}
\delta = \delta_{conv} + \frac{1}{2} (\delta_{os} - \delta_{conv}) \Bigg[1-\textrm{tanh}\, \Bigg(\frac{r - r_{tran}}{d_{tran}}\Bigg) \Bigg] \,,
\end{eqnarray}
 
\begin{eqnarray}
\delta_{conv} = \delta_{top}\, \textrm{exp}\, \Bigg(\frac{r - r_{max}}{d_{top}}\Bigg) + \delta_{cz}\, \frac{r-r_{sub}}{r_{max}-r_{sub}} \,,
\end{eqnarray}

\noindent where $\delta_{top}$, $\delta_{cz}$, and $\delta_{os}$ represent the values of superadiabaticity at the top of the domain ($r=r_{max}$), in the bulk of the convection zone, and in the overshoot region. The terms $r_{sub}$ and $r_{tran}$ denote the radius at which the stratification within the convection zone turns weakly subadiabatic and the radius of transition toward stronger subadiabatic stratification, respectively. The steepness of the transition towards larger superadiabaticities at the top of the domain and towards the overshoot regions are determined by $d_{top}$ and $d_{tran}$.

We assume that the diffusivities only depend on the radial coordinate and turbulent viscosity and thermal conductivity are given as:

\begin{eqnarray}
\nu_t = \frac{\nu_0}{2} \Bigg[1+\textrm{tanh}\,\Bigg(\frac{r-r_{tran}+\Delta}{d_{\kappa\nu}}\Bigg)\Bigg] f_c(r) \,,
\end{eqnarray}

\begin{eqnarray}
\kappa_t = \frac{\kappa_0}{2} \Bigg[1+\textrm{tanh}\,\Bigg(\frac{r-r_{tran}+\Delta}{d_{\kappa\nu}}\Bigg)\Bigg] f_c(r) \,,
\end{eqnarray}

\begin{eqnarray}
f_c(r) = \frac{1}{2} \Bigg[1+\textrm{tanh}\,\Bigg(\frac{r-r_{bc}}{d_{bc}}\Bigg)\Bigg] \,,
\end{eqnarray}

\begin{eqnarray}
\Delta = d_{\kappa\nu}\, \textrm{tanh}^{-1}\,(2\alpha_{\kappa\nu}-1) \,,
\end{eqnarray}

The radial and latitudinal angular momentum fluxes are proportional to the off-diagonal components of the correlation tensor in spherical coordinates, $Q_{r\phi}$ and $Q_{\theta\phi}$ and they are finite and parametrised as:

\begin{eqnarray}
Q_{r\phi}^{\Lambda} = \nu_{t}\, \Omega\, V\,\textrm{sin}\theta\,,
\end{eqnarray}

\begin{eqnarray}
Q_{\theta\phi}^{\Lambda} = \nu_{t}\, \Omega\, H\,\textrm{cos}\theta\,,
\end{eqnarray}

\noindent $V$ and $H$ are the normalised vertical and horizontal fluxes:

\begin{eqnarray}
V =  V^{(0)} (\Omega^{*}) - H^{(1)} (\Omega^{*})\,\textrm{cos}^2\theta \,,
\end{eqnarray}

\begin{eqnarray}
H =  H^{(1)} (\Omega^{*})\,\textrm{sin}^2\theta \,,
\end{eqnarray}

\begin{eqnarray}
V^{(0)}= \Bigg(\frac{\ell_{corr}}{H_\rho}\Bigg)^{1/2} \big(J_0 (\Omega^{*}) + aI_0\,(\Omega^{*})\big) \,,
\end{eqnarray}

\begin{eqnarray}
H^{(1)}= \Bigg(\frac{\ell_{corr}}{H_\rho}\Bigg)^{1/2} \big(J_1 (\Omega^{*}) + aI_1\,(\Omega^{*})\big) \,,
\end{eqnarray}

\noindent with

\begin{eqnarray}
J_0(\Omega^{*})= \frac{1}{2{\Omega^{*}}^{4}} \Bigg(9-\frac{2{\Omega^{*}}^{2}}{1+{\Omega^{*}}^{2}}-\frac{{\Omega^{*}}^{2}+9}{\Omega^{*}}\,\textrm{arctan}\Omega^{*}\Bigg) \,,
\end{eqnarray}

\begin{eqnarray}
J_1(\Omega^{*}) &=& \frac{1}{2{\Omega^{*}}^{4}} \Bigg(45+{\Omega^{*}}^{2} - \frac{4{\Omega^{*}}^{2}}{1+{\Omega^{*}}^{2}} \\ \nonumber
&&+\frac{{\Omega^{*}}^{4}-12{\Omega^{*}}^{2}-45}{\Omega^{*}}\,\textrm{arctan}\Omega^{*}\Bigg) \,,
\end{eqnarray}

\noindent and

\begin{eqnarray}
I_0(\Omega^{*}) &=& \frac{1}{4{\Omega^{*}}^{4}} \\ \nonumber
&&\times \Bigg(-19-\frac{5}{1+{\Omega^{*}}^{2}}+\frac{3{\Omega^{*}}^{2}+24}{\Omega^{*}}\,\textrm{arctan}\Omega^{*}\Bigg) \,,
\end{eqnarray}

\begin{eqnarray}
I_1(\Omega^{*}) &=& \frac{3}{4{\Omega^{*}}^{4}} \\ \nonumber
&&\times \Bigg(-15-\frac{2{\Omega^{*}}^{2}}{1+{\Omega^{*}}^{2}}+\frac{3{\Omega^{*}}^{2}+15}{\Omega^{*}}\,\textrm{arctan}\Omega^{*}\Bigg) \,,
\end{eqnarray}

\noindent where $\Omega^{*}$ denotes the Coriolis number.

The mean field differential rotation and meridional circulation model described above is then coupled with the axisymmetric mean field dynamo equations. The computed magnetic field is allowed to feed back on differential rotation and meridional circulation via the Lorentz force. The vector potential is introduced in the induction equation to satisfy the constrain $\nabla \cdot \boldsymbol{B} = 0$. The coupled equations are:

\begin{eqnarray}
\frac{\partial \varv_{r}}{\partial t} &=& - \varv_{r} \frac{\partial \varv_r}{\partial r} - \frac{\varv_\theta}{r} \frac{\partial \varv_r}{\partial \theta} + \frac{\varv^2_\theta}{r} - \frac{\partial}{\partial r}\frac{p_{tot}}{\varrho_0}+\frac{p_{mag}}{\gamma\,p_0}\,g \\ \nonumber
&&+ \frac{s_1}{\gamma}\,g + (2 \Omega_0 \Omega_1 + \Omega_1^2) r\, \textrm{sin}^2\theta\,+\frac{1}{\varrho_0}(F_r^\nu + F_r^B)  \,,
\end{eqnarray}

\begin{eqnarray}
\frac{\partial \varv_{\theta}}{\partial t} &=& - \varv_{r} \frac{\partial \varv_\theta}{\partial r} - \frac{\varv_\theta}{r} \frac{\partial \varv_\theta}{\partial \theta} - \frac{\varv_r \, \varv_\theta}{r} -  \frac{1}{r} \frac{\partial}{\partial \theta}\frac{p_{tot}}{\varrho_0}  \\ \nonumber
&&+ (2 \Omega_0 \Omega_1 + \Omega_1^2)\, r\, \textrm{sin}\theta\, \textrm{cos}\theta + \frac{1}{\varrho_0} (F_\theta^\nu + F_\theta^B)  \,,
\end{eqnarray}

\begin{eqnarray}
\frac{\partial \Omega_{1}}{\partial t} &=& - \frac{\varv_r}{r^2}  \frac{\partial}{\partial r} \, \bigg[r^2 (\Omega_0+ \Omega_1)\bigg] \\ \nonumber
&&- \frac{\varv_\theta^2}{r\,\textrm{sin}^2\theta} \frac{\partial}{\partial \theta}\bigg[\textrm{sin}^2\theta \,(\Omega_0 + \Omega_1)\bigg] + \frac{1}{\varrho_o\,r\,\textrm{sin}\theta} (F_\phi^\nu + F_\phi^B)\,,
\end{eqnarray}

\begin{eqnarray}
\frac{\partial s_1}{\partial t} &=& -\varv_r\, \frac{\partial s_1}{\partial r} - \frac{\varv_\theta}{r} \frac{\partial s_1}{\partial \theta} + \varv_r \frac{\gamma \delta}{H_p} + \frac{\gamma - 1}{p_0} Q \\ \nonumber
&& + \frac{1}{\varrho_0 T_0}\,\nabla\cdot(\kappa_t\,\varrho_0\,T_0\,\nabla\,s_1)+\frac{\gamma-1}{p_0}\eta_t(\nabla\times\boldsymbol{B})^2\,,
\end{eqnarray}

\begin{eqnarray}
\frac{\partial B_\phi}{\partial t} &=& - \frac{1}{r} \frac{\partial}{\partial r}(r\,\varv_r B_\phi) - \frac{1}{r}\frac{\partial}{\partial\theta}(\varv_\theta B_\phi) + r\,\textrm{sin}\theta B_r \frac{\partial \Omega_1}{\partial r} \\ \nonumber
&&+\textrm{sin}\theta B_\theta \frac{\partial \Omega_1}{\partial \theta} + \eta_t \Bigg[\Delta-\frac{1}{(r\textrm{sin}\theta)^2}\Bigg]B_\phi \\ \nonumber
&&+\frac{1}{r} \frac{\partial \eta_t}{\partial r} \frac{\partial}{\partial r}(r\,B_\phi) + \frac{1}{r^2} \frac{\partial \eta_t}{\partial \theta} \frac{1}{\textrm{sin}\theta} \frac{\partial}{\partial \theta}(\textrm{sin}\theta B_\phi) \,,
\end{eqnarray}

\begin{eqnarray}
\frac{\partial A}{\partial t} &=& -\frac{\varv_r}{r} \frac{\partial}{\partial r}(r\,A) - \frac{\varv_\theta}{r\textrm{sin}\theta} \frac{\partial }{\partial \theta}(\textrm{sin}\theta\,A) \\ \nonumber
&&+ \eta_t \Bigg[\Delta-\frac{1}{(r\textrm{sin}\theta)^2}\Bigg]\,A+S(r,\theta,B_{\phi}) \,,
\end{eqnarray}

In the above equations, $\eta_t$ denotes turbulent magnetic diffusivity, which is given as

\begin{eqnarray}
\eta_t=\eta_c+f_c(r)[\eta_{bc}-\eta_c+f_{cz}(r)(\eta_{cz}-\eta_{bc})] \,,
\end{eqnarray}

\begin{eqnarray}
f_{cz}(r)=\frac{1}{2} \Bigg[1+\textrm{tanh}\Bigg(\frac{r-r_{cz}}{d_{cz}}\Bigg)\Bigg] \,,
\end{eqnarray}

\begin{eqnarray}
f_{c}(r)=\frac{1}{2} \Bigg[1+\textrm{tanh}\Bigg(\frac{r-r_{bc}}{d_{bc}}\Bigg)\Bigg] \,,
\end{eqnarray}

The source term $S(r,\theta,B_{\phi})$ in the induction equation is given as:

\begin{eqnarray}
S(r,\theta,B_{\phi})=\alpha_{0}\overline{B}_{\phi, bc}(\theta)f_{\alpha}(r)g_{\alpha}(\theta)
\end{eqnarray}

\noindent where 

\begin{eqnarray}
\overline{B}_{\phi, bc}(\theta)=\int_{r_{min}}^{r_{max}} dr h(r) B_{\phi} (r, \theta)
\end{eqnarray}

For for the nonlocal the BL-effect:

\begin{eqnarray}
f_{\alpha}(r)=max\bigg[0, 1-\frac{(r-r_{max})^{2}}{d_{\alpha}^{2}}\bigg]
\end{eqnarray}

\begin{eqnarray}
g_{\alpha}(r)=\frac{(\textrm{sin}\theta)^{4} \, \textrm{cos}\theta}{max\big[(\textrm{sin}\theta)^{4} \, \textrm{cos}\theta\big]}
\end{eqnarray}

As for the turbulent $\alpha$-dynamo operating in the bottom half of the convection zone, the source term becomes:

\begin{eqnarray}
f_{\alpha}(r)=\frac{1}{4}\bigg[1+erf\bigg(\frac{r-r_{2}}{d_{2}}\bigg)\bigg]\bigg[1-erf\bigg(\frac{r-r_{3}}{d_{3}}\bigg)\bigg]
\end{eqnarray}

\begin{eqnarray}
g_{\alpha}(r)=\frac{\textrm{sin}\theta \, \textrm{cos}\theta}{max\big[\textrm{sin}\theta \, \textrm{cos}\theta\big]}
\end{eqnarray}

\end{appendix}
\end{document}